\documentclass[preprint,onecolumn,showpacs,preprintnumbers,amsmath,amssymb]{revtex4}
\usepackage{graphicx}
\usepackage{dcolumn}
\usepackage{bm}

\newcommand{\bes}{\begin{eqnarray}}
\newcommand{\ees}{\end{eqnarray}}
\newcommand{\be}{\begin{equation}}
\newcommand{\ee}{\end{equation}}

\begin{document}
\title{Ag-Cu alloy surfaces in an oxidizing environment: a first-principles study}
\author{Simone Piccinin, Catherine Stampfl}
\affiliation {School of Physics, The University of Sydney, Sydney, New South Wales 2006, Australia}
\author{Matthias Scheffler}
\affiliation{Fritz-Haber-Institut der Max-Planck-Gesellschaft, Faradayweg 4-6, D-14195 Berlin, Germany}
\date{\today}

\begin{abstract}

Recent experiments on model catalysts have shown that Ag-Cu alloys have improved selectivity with respect to pure silver for ethylene epoxidation. In this paper we review our first-principles investigations on the (111) surface of this alloy and present new findings on other low index surfaces. We find that, for every surface orientation, the presence of oxygen leads to copper segregation to the surface. Considering the alloy to be in equilibrium with an oxygen atmosphere and accounting for the effect of temperature and pressure, we compute the surface free energy and study the stability of several surface structures. Investigating the dependence of the surface free energy on the surface composition, we construct the phase diagram of the alloy for every surface orientation. Around the temperature, pressure and composition of interest for practical applications, we find that a limited number of structures can be present, including a thin layer of copper oxide on top of the silver surface and copper-free structures. Different surface orientations show a very similar behavior and in particular a single layer with CuO stoichiometry, significantly distorted when compared to a layer of bulk CuO, has a wide range of stability for all orientations. Our results are consistent with, and help explain, recent experimental measurements.
\end{abstract}

\maketitle

\section{Introduction}

Processes such as heterogeneous catalysis and corrosion are dominated by chemical reactions that happen on the surfaces of materials~\cite{StampflSS2002}. These processes involve length and time scales that span several orders of magnitude: on one hand, molecular reactions occur on a sub-nanometer length and picosecond time scale, while, on the other hand, for surface phase transitions, catalysis, and corrosion the relevant time scale is of the order of microseconds or even seconds. Clearly this calls for a multiscale approach if one aims at describing, e.g., the full catalytic cycle or the full corrosion problem.~\cite{HandbookMM} 

Our present study aims at obtaining a detailed knowledge of the atomic and molecular processes that underlie such phenomena. We are therefore interested in providing an atomistic description of the transformations that take place at surfaces, in particular in the case of heterogeneous catalysis. The study of the formation or the breaking of a chemical bond, in any reaction, requires a quantum-mechanical description of the electrons. This calls for a first-principles description that does not rely on empirical parameters but rather just on the knowledge of the types of atoms present in the materials or molecules involved in the processes. It also requires the consideration of the quantum-mechanical kinetic energy operator in order to account for the shell structure of atoms, molecular orbitals, and chemical bonds. Among the electronic structure theories that have been developed to tackle such problems, density functional theory (DFT)~\cite{HK64, KS65} has emerged as the most successful approach, given the very favorable ratio between accuracy and computational cost that this theory provides. 

As a first step toward modeling heterogeneous catalysis, an atomic description of the structure of the surfaces is required.~\cite{StampflCT2005} For decades surface science techniques have been employed to study surface structures and transformations in ultra high vacuum (UHV) conditions and the knowledge thus obtained has been applied to ``real life'' catalytic processes that often take place at high temperature and high pressure regimes. While, in many cases, this approach has been extremely successful (see e.g.~Ref.~\onlinecite{ErtlACIE2008}), the pressure gap between surface science experiments and realistic applications inhibits, in several cases, a direct transfer: This is because the structure and composition of the surface can drastically change when going from UHV to realistic conditions~\cite{StampflSS2002,HandbookMM}. A typical system that undergoes such a transformation is ruthenium: the catalytic activity toward CO oxidation under reactant pressures of the order of atmospheres has been in fact attributed to the formation of an RuO$_2$ oxide overlayer on Ru, rather than pure Ru (see Ref.~\cite{ReuterCPL02} and references therein). Moreover, in this system the highest catalytic activity is reached in regions of the phase diagram corresponding to boundaries between different stable surface oxide structures, and the surface structure and stoichiometry at the steady state of catalysis is very different to any thermodynamically stable surface. Experiments on CO oxidation at Pt(110) and Pd(100) conducted in high temperature and high pressure conditions~\cite{FrenkelCAT05} have also shown how, under steady-state conditions, thin oxides form at the surface and they continuously evolve with time. These recent results have therefore highlighted both the importance of the environment and the non-equilibrium nature of catalysis: The catalyst surface cannot be viewed as a static object but rather as a system that, at the atomistic level, continuously evolves and fluctuates.

It is therefore clear that the effects of temperature and pressure must be properly accounted for in order to obtain a reliable description of the surface structure and the processes happening on it.
To this end, first-principles calculations, which combine accurate electronic structure methods with equilibrium thermodynamics, have been successfully employed to predict the stability of surface structures of several systems under high temperature and high pressure conditions.~\cite{ReuterPRL03} Moreover, such methods have been included in a multiscale approach that employs kinetic Monte Carlo to simulate the dynamics of the full steady-state catalysis. In some cases the agreement with experiments can be quantitative, allowing an unprecedented level of insight into the concerted actions of a dynamical process such as heterogeneous catalysis.~\cite{ReuterPRL04} Hence, combining accurate electronic structure methods and non-equilibrium statistical mechanics techniques is now emerging as a possible way to tackle the multiscale nature of such processes.

Furthermore, research carried out in the last few years on oxidation reactions on transition metal catalysts have shown how the catalyst surface can present a number of possible structures, rather than a single low energy one. This suggests a scenario in which the catalyst dynamically evolves in time, fluctuating between structures with similar energetics. This picture has emerged, for example, in recent studies of CO oxidation at RuO$_2$(110)~\cite{ReuterPRL04} and Pd(100)~\cite{RogalPRL07}: In both cases the highest catalytic activity is reached in regions of the phase diagram corresponding to boundaries between different stable surface oxide structures. Experimentally, it has been found through scanning tunneling microscopy (STM) measurements conducted under realistic catalytic conditions for CO oxidation at Pt(110) and Pd(100)~\cite{FrenkelCAT05} that, under steady-state catalysis, the thin oxides present at the surface continuously evolve with time. These recent results have therefore stressed how the catalyst surface cannot be viewed as a static object but rather as a ``living'' system that, at the atomistic level, continuously evolves due to various processes such as adsorption, desorption, association, dissociation, diffusion and to the competition between structures with similar energetics.   

The use of bimetallic catalysts has been the focus of much work in the field of heterogeneous catalysis,~\cite{SinfeltBOOK} since the catalytic activity and selectivity of a metal can be modified substantially by alloying with another metal. For example, geometrical and electronic effects obtained by varying the alloy composition may play a crucial role in determining the properties of the catalyst.~\cite{Liu01} One appealing aspect of bimetallic alloys is the possibility of rationally designing the catalytic properties of the material by changing the alloy elements and composition. To this end, first-principles calculations have been shown to be a potentially useful tool for screening among pools of possible alloys and for extracting trends and therefore gaining insights into the functioning of the alloy.~\cite{Greeley04, Greeley06}. In view of the above discussion, though, the effects of high pressure and temperature and the role of the dynamical evolution of the catalyst must be carefully accounted for, in order for the theoretical simulations to have predictive power.

A catalytic process in which the use of bimetallic catalysts could provide a significant benefit is ethylene epoxidation, which is one of the most important processes in the chemical industry. Here, an ethylene molecule (C$_2$H$_4$) is oxidized to ethylene oxide (C$_2$H$_4$O), which is used in the synthesis of ethylene glycol, polyester fabrics, surfactants, and detergents. The catalyst for this reaction must promote the formation of ethylene oxide and at the same time inhibit the competing reaction, total oxidation to CO$_2$, which is thermodynamically more favourable. The selectivity toward the formation of ethylene oxide is therefore a key factor. In industrial applications, the catalyst used for this important chemical reaction is silver, and the reaction is carried out at atmospheric pressure and temperatures $T$=500-600 K.~\cite{Santen87} Both the selective (formation of ethylene oxide) and unselective  (total oxidation to CO$_2$, which occurs via the formation of acetaldehyde) paths have been found to originate from a common intermediate, an oxametallacycle, in which the molecule is bound to both a chemisorbed oxygen atom and a surface atom.~\cite{LinicJACS2002,LinicJACS2003} This intermediate can react to form either ethylene oxide or acetaldehyde with similar activation barriers. On the basis of both first-principles calculations~\cite{LinicJC2004} and experiments~\cite{JankJC2005} it has been suggested that if an Ag-Cu alloy, rather than pure Ag, is used as a catalyst, the selectivity toward ethylene oxide will be improved. 

In several recent works the mechanism behind the enhanced selectivity due to alloying with copper has been addressed.~\cite{LinicJC2004,TorresJACS2005,KokaljJC2008} The composition and structure of the catalyst surface, though, has not been thoroughly investigated. Given the discussion presented above, it is clear that the mechanism of this catalytic reaction can be strongly influenced by the atomistic details of the surface. In particular, given the conditions of operation of the Ag-Cu catalyst in realistic applications, the formation of surface oxides cannot be ruled out. It has been established that silver can form several surface oxide structures~\cite{MichaelidesJVCT05,SchnadtPRL2006,SchmidPRL2006} and from the phase diagram of the Cu-O system, considering the temperature and oxygen pressure used in practical applications, but neglecting the effect of ethylene, the formation of CuO is expected, therefore suggesting that copper might also oxidate.

In this paper we review our work on the first-principle investigations of the structure of the Ag-Cu(111) surface under varying oxygen pressure and temperature conditions, and we present new results on other low-index surfaces. In our calculations we consider the alloy surface to be in thermodynamic equilibrium with an atmosphere of pure oxygen. The effect of the presence of other reactants such as ethylene is not investigated in this work. Under the conditions of steady-state catalysis it is therefore likely that the surface of the catalyst might be modified with respect to the stable structures found in a pure oxygen atmosphere. Hence our work can be thought of as a first step in gaining insight into the possible structures of the alloy, an essential prerequisite for modeling the full catalytic process.     
The paper is organized as follows: In Sec.~\ref{Method} we briefly review the theoretical background relevant for this work, present the definitions of the quantities that will be used throughout the paper and discuss the approximations and assumptions that underpin our methodology. In Sec.~\ref{Results} we report our results and discuss their relevance. Section~\ref{Concl} summarizes the main findings.   
\section{Calculation Method}
\label{Method}

The density functional theory (DFT) calculations presented in this work are performed using the generalized gradient approximation (GGA) of Perdew-Burke-Ernzerhof (PBE)~\cite{PBE} for the exchange and correlation functional. The electron-ion interactions are computed using ultrasoft pseudopotentials,~\cite{USPP,USespresso} including scalar relativistic effects, and the energy cutoff for the plane wave expansion is 27 Ry (200 Ry for the charge density cutoff). The Brillouin zone is sampled using a Monkhorst-Pack mesh,~\cite{Monkhorst-Pack} broadening the Fermi surface according to the Marzari-Vanderbilt cold-smearing technique,~\cite{MVcold} using a smearing parameter of 0.03 Ry. A 12$\times$12$\times$1 \textbf{k}-point mesh is used for the (1 $\times$ 1) surface unit cell of the (111) and (100) surfaces, while a 12$\times$9$\times$1 mesh is used for the (1 $\times$ 1) surface unit cell of the (110) surface.  For larger unit cells, the \textbf{k}-point mesh is scaled accordingly to give us identical \textbf{k}-point sampling. All the calculations are performed using the PWscf code contained in the Quantum-ESPRESSO package.~\cite{espresso} We use a slab geometry to model the surfaces, using 4 Ag atom layers for the (111) surface, 6 for the (100) and 6 for the (110), on top of which the adsorption structures are created, on one side on the slab only. For large unit cells, in particular for the (4 $\times$ 4) structures on the (111) surface, we reduced the number of Ag layers to three. The bottom two layers are fixed to the bulk positions, while all other atoms are allowed to relax until the forces are less than 0.001 Ry/Bohr (0.025 eV/\AA). A 12 \AA~vacuum layer is used, which is found to be sufficient to ensure negligible coupling between periodic replicas of the slab. Further details of the calculations presented in this work can be found in one of our previous publications~\cite{PiccininPRB2008}.      

To help the discussion presented in the following section, we now define several quantities. Considering the adsorption of atomic oxygen on a clean Ag surface, we define the average binding energy per oxygen atom as
\be
E^{{\rm O/Ag}}_b = -\frac{1}{N_{\rm O}}[E^{{\rm O/Ag}} - (E^{{\rm slab}} + N_{\rm O} E^{\rm O})]\quad,
\label{eq_eb}
\ee
where $N_{\rm O}$ is the number of oxygen atoms in the unit cell, $E^{{\rm O/Ag}}$, $E^{\rm slab}$ and $E^{\rm O}$ are the total energies of the adsorption system, the Ag slab and half the oxygen molecule, i.e. $E^{\rm O} = 1/2E^{\rm total}_{{\rm O}_2}$. With this definition, a positive binding energy indicates that the adsorption of an oxygen atom is exothermic with respect to oxygen in molecular gas phase form. 
Here we point out that the significant error ($\sim$ 0.54 eV) in the oxygen molecule binding energy introduced by the GGA-PBE approximation ($1/2E^{\rm O_2}_b = 3.10$ eV, while the experimental value is 2.56 eV~\cite{HuberBOOK}) introduces a large error bar in the determination of the oxygen chemical potential. This error, however, is partially be compensated by the analogous GGA-PBE error in the description of oxygen chemisorbed on the metal surfaces.


To account for the effects of temperature ($T$) and pressure ($p$) we employ ``{\it ab initio} atomistic thermodynamics'',~\cite{WeinerMSF86, WangPRL2000, ReuterPRB2002, LiPRB2003, StampflCT2005} which allows the determination of the lowest energy structures as a function of $T$ and $p$. The surface is considered to be in contact with an oxygen atmosphere that acts as a reservoir, therefore exchanging oxygen atoms with the surface without changing its temperature and pressure (i.e. its chemical potential).

The change in Gibbs free energy is calculated as  
\be
\Delta G(\mu_{\rm O}) = \frac{1}{A}(G^{\rm O/Ag} - G^{\rm{slab}}-\Delta N_{\rm{Ag}}\mu_{\rm{Ag}}-N_{\rm{O}}\mu_{\rm{O}}) \quad ,
\label{eq_gamma1}
\ee
where $G^{\rm O/Ag}$ is the free energy of the adsorbate/substrate structure, $G^{\rm slab}$ is the free energy of the clean Ag slab, $\Delta N_{\rm{Ag}}$ is the difference in the number of Ag atoms between the adsorption system and the clean Ag slab, $\mu_{\rm{Ag}}$ and $\mu_{\rm{O}}$ are the atom chemical potentials of Ag and O. The change in Gibbs free energy is normalized by the surface area $A$ to allow comparisons between structures with different unit cells; we will refer to this quantity as the ``change in Gibbs surface free energy of adsorption'', or simply ``surface free energy''. The chemical potential of Ag is taken to be that of an Ag atom in bulk Ag, therefore assuming that the slab is in equilibrium with the bulk, that acts as the silver reservoir. The oxygen chemical potential depends on temperature and pressure according to~\cite{ReuterRuO2}
\be
\mu_{\rm O}(T,p) = \frac{1}{2}[ E^{\rm total}_{\rm O_2}(T, p^{\rm 0}) + \tilde\mu_{\rm O_2}(T,p^{\rm 0}) + k_BTln\left( \dfrac{p_{\rm O_2}}{p^{\rm 0}}\right)] \quad . 
\label{EqOxymu} 
\ee 
Here $p^{\rm 0}$ is the standard pressure and $\tilde\mu_{\rm{O}_2}(T,p^0)$ is the chemical potential at the standard pressure, which can obtained either from thermochemical tables~\cite{thermotables} (the choice made in this study) or directly computed.

While the free energy should include terms that depend on the vibrational modes of the system, as well as configurational disorder, in our treatment we neglect such contributions. Especially when comparing systems with different stoichiometries, such terms can be non-negligible. For our system, however, through a simple estimate based on the Einstein model and approximating the phonon density of states by just one characteristic frequency for oxygen, we have shown~\cite{PiccininPRB2008} that neglecting the vibrational contributions to the free energy does not alter the main findings of our work. Also the configurational entropy has been found, through a simple estimate, not to contribute significantly to the free energy.~\cite{PiccininPRB2008}  
Hence, within this model, the only term that depends on $T$ and $p$ is the oxygen chemical potential. As a consequence, the free energies $G^{\rm O/Ag}$ and $G^{\rm slab}$ are identified as the total energies $E^{\rm O/Ag}$ and $E^{\rm slab}$. 

If we now consider the presence of copper impurities in the silver surface, we need to modify the definition of surface free energy. As we will show later in the paper, we deal with structures in which an overlayer containing O, Ag and Cu is adsorbed on a clean Ag surface. If we consider Cu to be in equilibrium with a bulk Cu reservoir, the definition of surface free energy becomes:
\be
\Delta G(\mu_{\rm O}) = \frac{1}{A}(G^{\rm O/Cu/Ag} - G^{\rm slab}-\Delta N_{\rm{Ag}}\mu_{\rm{Ag}} - N_{\rm{Cu}}\mu_{\rm{Cu}} -N_{\rm{O}}\mu_{\rm{O}}) \quad ,
\label{eq_gamma2}
\ee
where $G^{\rm O/Cu/Ag}$ is the free energy of the adsorption system, $N_{\rm Cu}$ is the number of Cu atoms and $\mu_{\rm{Cu}}$ the copper chemical potential. Since in this work, as we will show in Sec.~\ref{Results}, we are interested in identifying the structures belonging to the convex hull of the free energy vs. copper content curve, which depends on the curvature of such curve, rather than the absolute surface free energy, the choice of the Cu chemical potential is arbitrary, since the surface free energy depends linearly on it (see Eq.~\ref{eq_gamma2}).
We now define the average oxygen binding energy in the whole structure as
\be
E^{{\rm O/Cu/Ag}}_b = -\frac{1}{N_{\rm O}}[E^{{\rm O/Cu/Ag}} - (E^{{\rm slab}} + 
\Delta N_{\rm{Ag}}\mu_{\rm{Ag}} + N_{\rm{Cu}}\mu_{\rm{Cu}} + N_{\rm O} E^{\rm O})]\quad.
\label{eq_eb1}
\ee
Using the above-mentioned approximations, we can express the change in Gibbs surface free energy in terms of the average oxygen binding energy:
\bes
\Delta G(\Delta\mu_{\rm O}) = &&-\frac{1}{A}(N_{\rm O} E_b^{\rm O/Cu/Ag} + N_{\rm{O}}\Delta\mu_{\rm{O}}) \quad ,
\label{eq_gamma3}
\ees
where the oxygen chemical potential is now referenced with respect to half an isolated O$_2$ molecule: $\Delta \mu_{\rm O} = \mu_{\rm O} - \frac{1}{2}E^{\rm total}_{\rm O_2}$.

In writing the change in free energy as in Eq.~(\ref{eq_gamma3}) we are neglegting the contribution to the free energy coming from the configurational entropy. It has been shown~\cite{PiccininPRB2008} that indeed this approximation introduces negligible effects for the range of temperatures we are interested in. 
\section{Results}
\label{Results}

\subsection{Oxygen-induced segregation of copper to the surface}

Due to their significant (13 $\%$) lattice mismatch, copper and silver are known to be almost completely
immiscible in the bulk at temperatures of interest for industrial applications.
In several cases, however, elements which are immiscible in the bulk have been found to form stable
two-dimensional alloys at the surface~\cite{TersoffPRL1995}. These include Na and K 
deposited on Al(111) and (001),~\cite{NeugebauerPRL93} Au on Ni(110),~\cite{NielsenPRL93} 
Ag on Pt(111)~\cite{RoderPRL93} and Sb on Ag(111).~\cite{OppoPRL93} 
Indeed, the presence of such a two-dimensional alloy has been invoked also for the Ag-Cu 
system in the work by Linic {\it et al.}~\cite{LinicJC2004}

\begin{figure}[htb]
\begin{minipage}[c]{0.49\textwidth}
\centering
\includegraphics[width=65mm, angle=270]{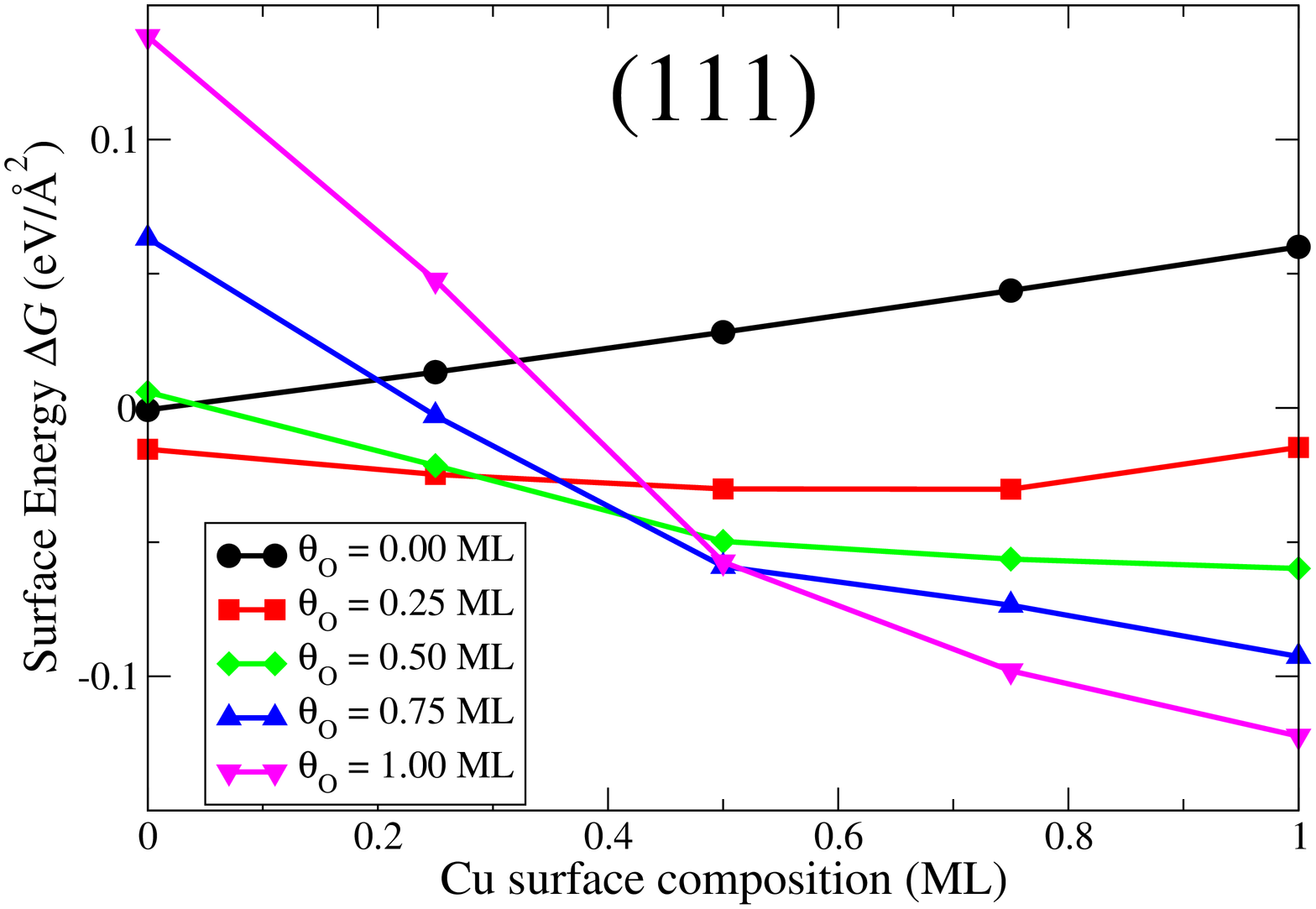}
\end{minipage}
\begin{minipage}[c]{0.49\textwidth}
\centering
\includegraphics[width=65mm, angle=270]{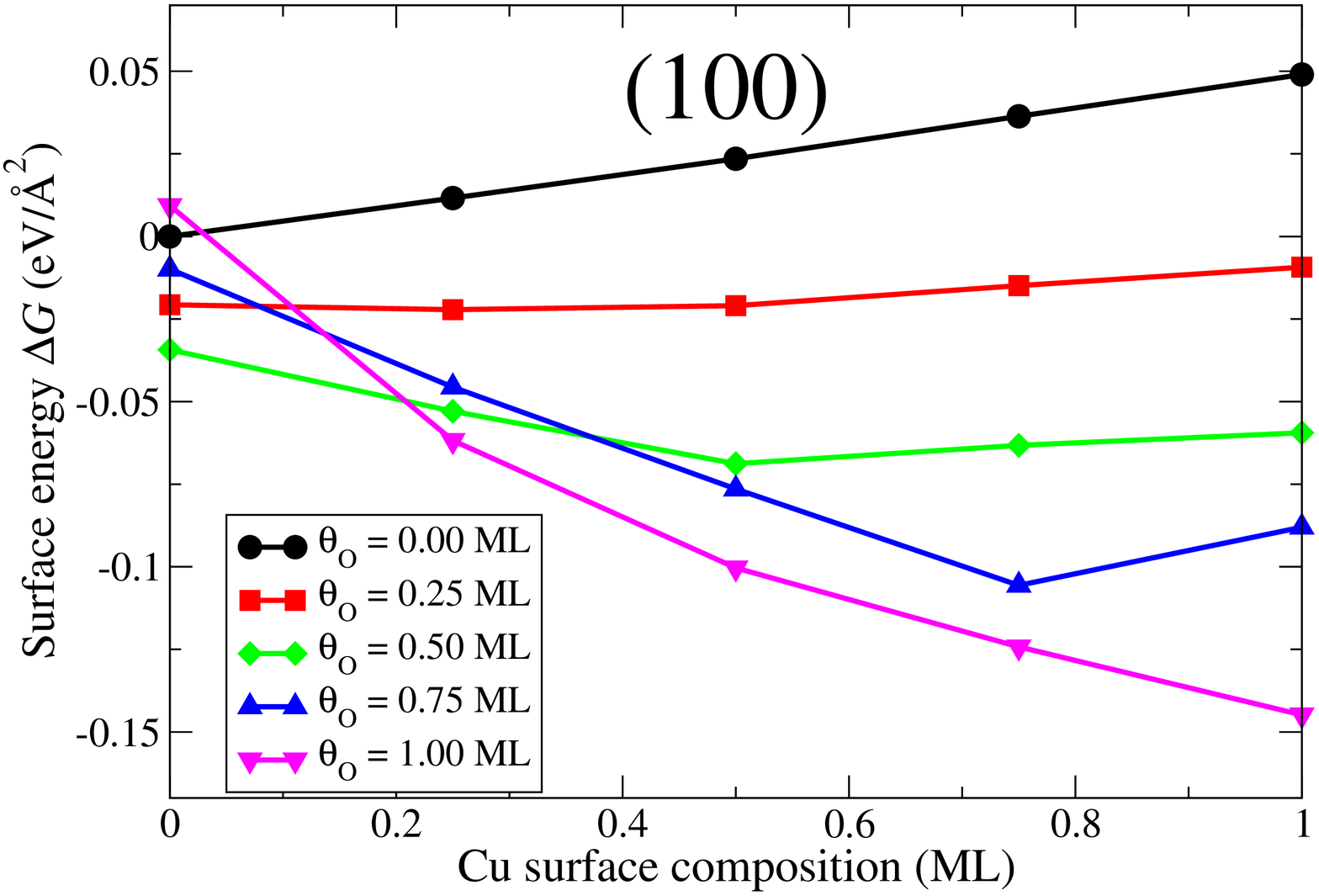}
\end{minipage}
\begin{minipage}[c]{1.0\textwidth}
\centering
\includegraphics[width=65mm, angle=270]{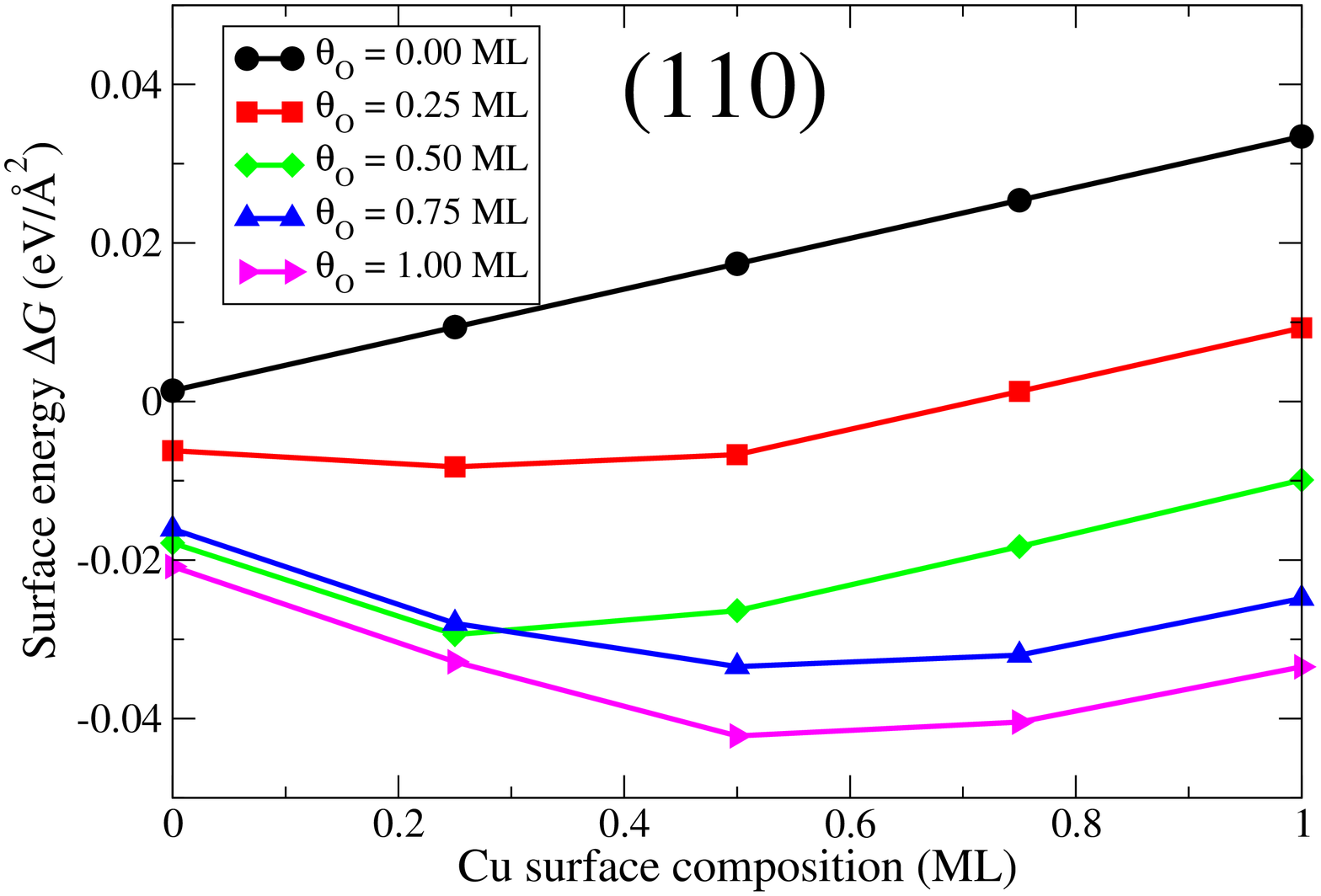}
\end{minipage}
\caption{\label{FigSE} (Color online) Change in Gibbs surface free energy at $T = 0$ K (cf. Eq.~\ref{eq_gamma3}) 
as a function of Cu concentration in the first layer of the surface. The oxygen coverages considered are indicated in the legend in monolayers (ML). The energy zero corresponds the surface energy of the clean Ag(111) surface. The top left panel refers to the (111) surface, the top right to the (100) surface and the bottom one to the (110) surface.}
\end{figure}

Considering the (111) surfaces of Ag and Cu, due to the fact the surface energy of silver is
considerably lower than the one of copper (experimental values: $\gamma^{\rm Ag(111)}= 0.078$ eV/\AA$^2$~\cite{BoerBOOK} and $\gamma^{\rm Cu(111)}= 0.114$ eV/\AA$^2$~\cite{Lindgren84}), it
is unlikely for copper impurities in silver to be exposed on the surface. We verified this
by computing the total energy of a (111) silver slab in which every fourth Ag surface atom is 
substituted with a Cu atom. By changing the layer of substitution, we found that the 
most favorable position for a Cu impurity is one layer below the surface. We also found that an 
increase in the copper content on the surface leads to an increase in surface energy. This is true for
all the three surface orientations considered here, i.e. (111), (100) and (110). The
results are shown in Fig.~\ref{FigSE} as the curves with black dots. The positive slope of the
curves indicate that it is unfavorable for Cu to migrate to the surface. The black curves
in Fig.~\ref{FigSE} have an almost negligible convexity, indicating a very small tendency to 
form a surface alloy. The configurational entropy contribution to the free energy, neglected
in the present approach, would further enhance the stability of an alloy at finite temperature. 
At T=600 K we estimate this term to contribute 5 meV/\AA$^2$ to the surface free energy.~\cite{PiccininPRB2008}
We therefore find that for this system, when Cu is confined to the first layer and in the absence 
of oxygen, there is a weak tendency to form a two-dimensional alloy. Similar results have been found 
for an Ag-Cu alloy on the Cu(100) surface.~\cite{NorskovSurfAll} 

Cu thin-film growth on Ag(111) in ultra high vacuum (UHV) has been recently studied experimentally.~\cite{MaurelSS2005,BocquetPRB2005} It was found that at room temperature, 
upon deposition of Cu on Ag(111), Cu forms islands that are encapsulated by one monolayer of Ag.
These findings agree with our theoretical results which show that the most favourable configuration 
for Cu is to sit one layer below the Ag surface layer.
     
When oxygen is introduced into the system, the picture changes completely:
We find that the presence of oxygen chemisorbed on the alloy surface has the remarkable effect of reversing the slope of the (black dots) curves for the surface energy versus Cu surface composition, shown in Fig.~\ref{FigSE}. For these calculations we consider an Ag-Cu alloy in the first layer of the Ag slab, with oxygen adsorbed in the most favorable adsorption sites. By fully relaxing the structure, we calculate the change in surface free energy at $T = 0$ K as a function of the Cu content in the first layer and the oxygen coverage. The results are shown in Fig.~\ref{FigSE} for the three surfaces considered in this work. We can see that for the (111) surface the presence of at least a quarter of a monolayer (ML) of oxygen induces Cu to segregate to the surface, i.e. it changes the slope of the curve from positive (for an oxygen coverage relative to the underlying Ag(111) lattice $\theta_{\rm O} < 0.25$ ML) to negative (for $\theta_{\rm O} \geqslant 0.25$ ML). The driving mechanism here is the strong affinity between oxygen and copper, which more than compensates the unfavorable surface energy of Cu with respect to Ag: The adsorption energy (cf. Eq.(\ref{eq_eb})) of oxygen at 0.25 ML coverage on Ag(111) and Cu(111) is 0.38 eV/atom and 1.57 eV/atom, respectively. We also see that in the absence of copper (i.e. the left end of the graphs in Fig.~\ref{FigSE}), the most stable structure is the one with $\theta_{\rm O} = 0.25$ ML, i.e. the lowest O coverage. At higher oxygen coverages the oxygen-oxygen repulsion overcomes the formation of O-Ag bonds, leading to an increase in surface energy with the oxygen coverage. At the opposite end of the graph, where a full monolayer of copper is present in the first layer of the slab, the surface energy decreases with the oxygen coverage: This is due to the contribution to the surface energy of the formation of strong O-Cu bonds. The (100) and (110) surfaces display a behavior qualitatively similar to the (111) surface, but the transition to a negative slope takes place at a higher oxygen coverage, 0.50 ML and 0.75 ML for the (100) and (110) surface respectively. 

A study similar to the one shown here has been recently presented for the Ag$_3$Pd alloy.~\cite{KitchinPRB2008} Also in that system the presence of oxygen has the effect of inverting the segregation profile: While oxygen poor conditions favour Ag-terminated surfaces, in oxygen rich conditions Pd is driven to the surface, due to the strength of the Pd-O bond compared to the Ag-O one. The Pt-Ru alloy displays analogous properties as well,~\cite{HanPRB2005} where the more noble Pt is exposed in absence of oxygen, while the strong Ru-O bond leads to Ru segregation at the surface in oxygen rich conditions. All these systems therefore display the same trend: In absence of oxygen the element with the lowest surface energy is exposed, while in the presence of oxygen the element with the higher affinity with oxygen segregates at the surface.   

\subsection{Stable structures on the Ag-Cu alloy surfaces}

Having established that in the presence of oxygen, for all the three surface orientations considered, there is a clear tendency for Cu to segregate to the surface, in the following we assume that Cu is confined to the surface and we investigate what are the most stable surface structures as a function of temperature and pressure. Given the fact that copper might oxidize under the typical conditions of oxygen pressure and temperature in which the catalyst is supposed to operate, we also consider surface structures that are derived from the corresponding bulk copper oxide geometries. Here we point out that the real catalyst operates in the presence not only of oxygen but also of ethylene, which acts as the reducing agent. Since in our calculations we do not consider the presence of ethylene, the stability of the structures we study here could be altered by accounting for the effects all the reagents.
  
We study, for each of the three surface orientations, three types of structures: ({\it i}) chemisorbed oxygen on the Ag-Cu alloy in the first layer of the Ag slab (as described in the previous section), ({\it ii}) structures derived from copper(I) oxide Cu$_2$O, whose structure can be visualized as trilayers of O-Cu-O piled up on top of each other and ({\it iii}) structures derived from copper(II) oxide CuO.

\begin{figure}[tb]
\begin{minipage}[c]{0.45\textwidth}
\centering
\includegraphics[width=70mm, angle=0]{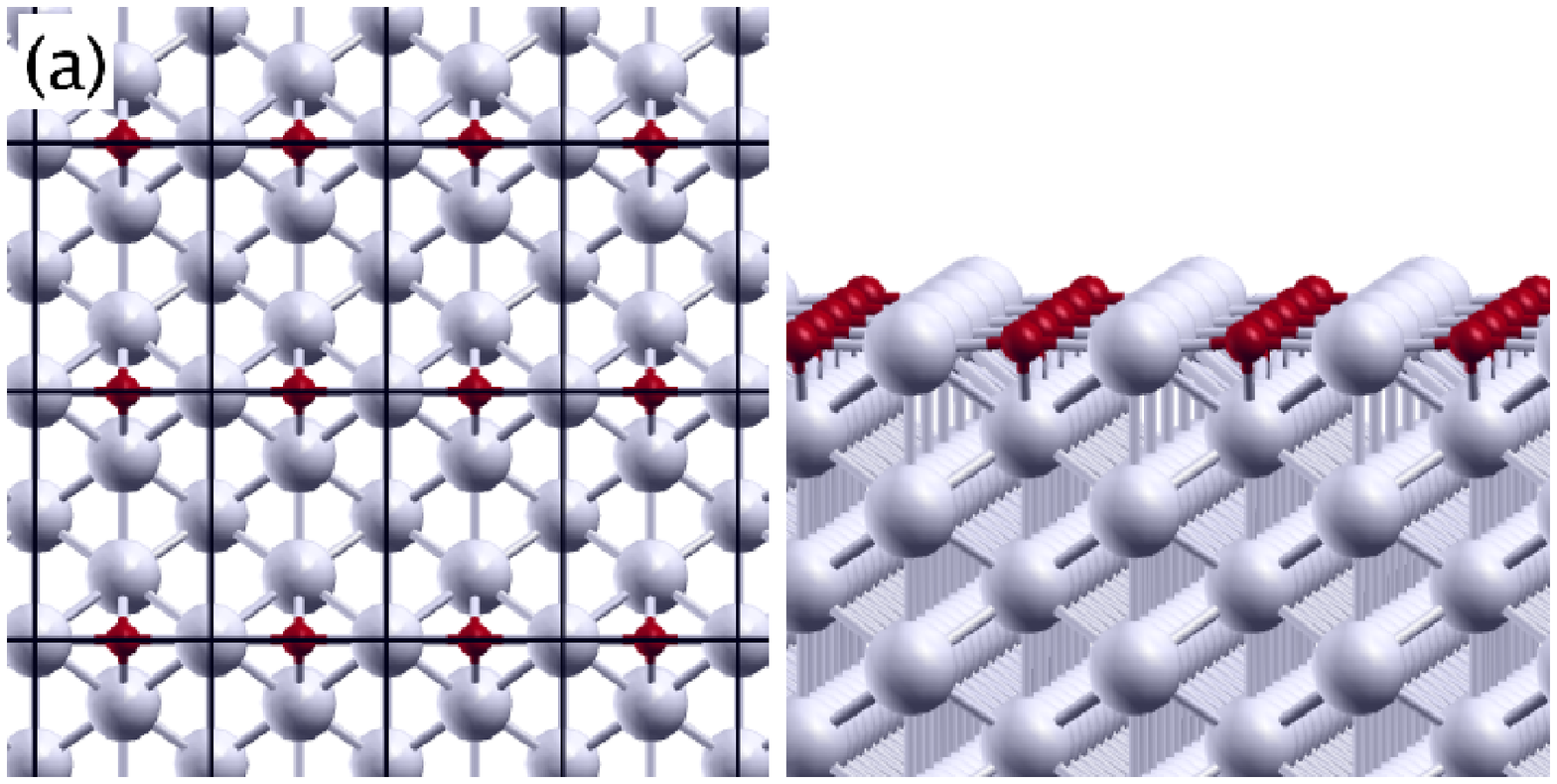}
\end{minipage}
\begin{minipage}[c]{0.45\textwidth}
\centering
\includegraphics[width=70mm, angle=0]{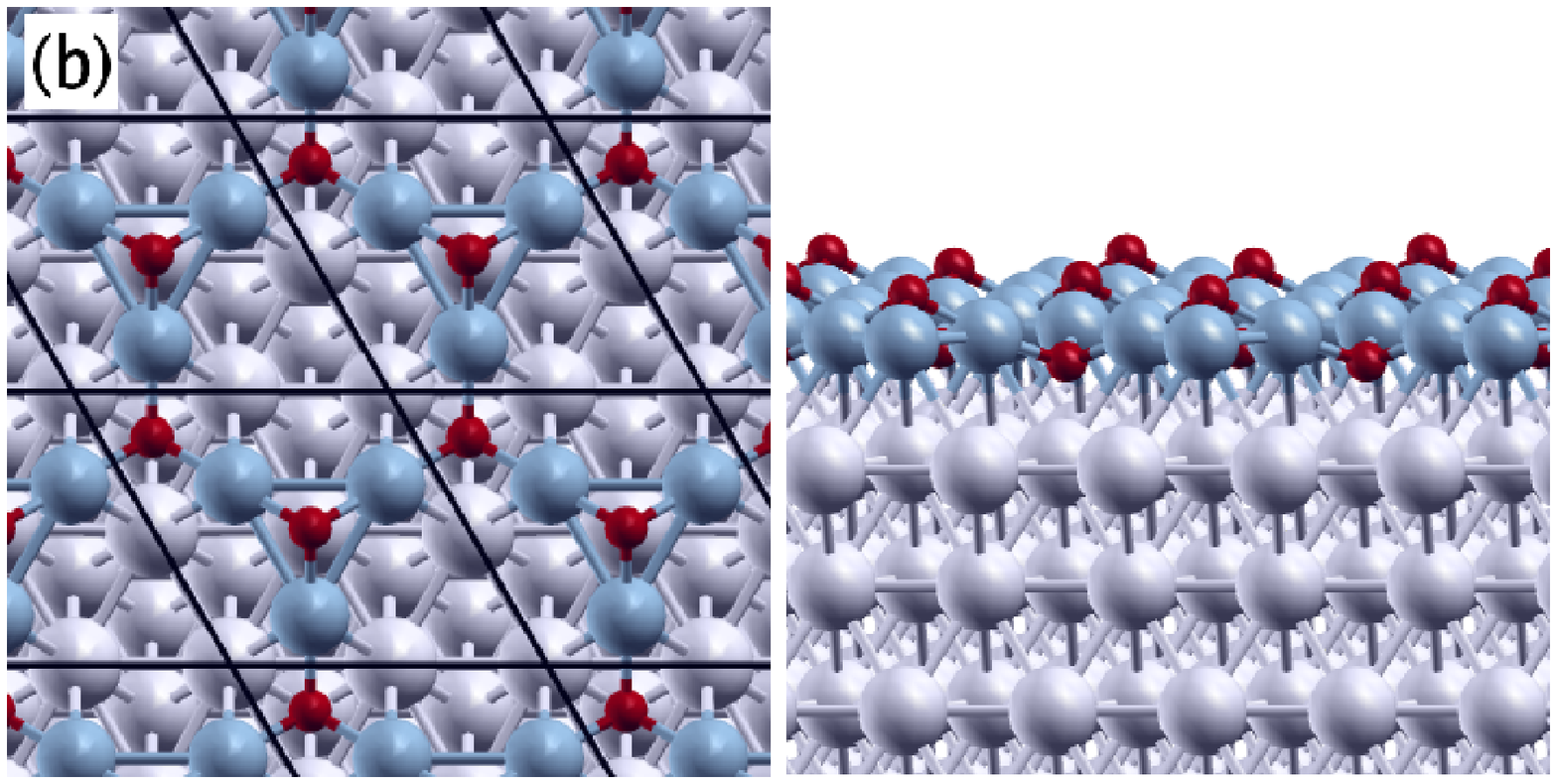}
\end{minipage}
\begin{minipage}[c]{0.45\textwidth}
\centering
\includegraphics[width=70mm, angle=0]{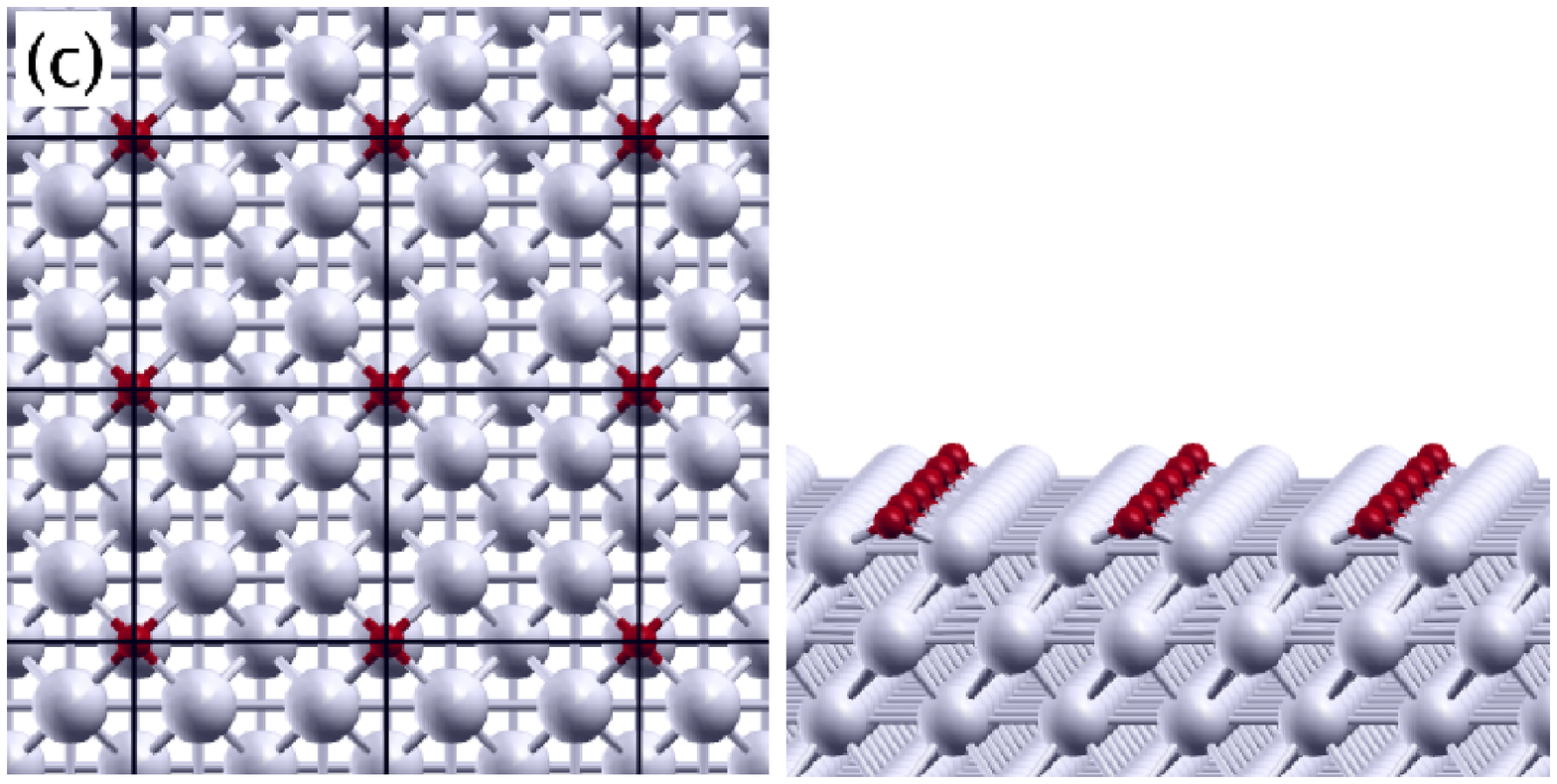}
\end{minipage}
\begin{minipage}[c]{0.45\textwidth}
\centering
\includegraphics[width=70mm, angle=0]{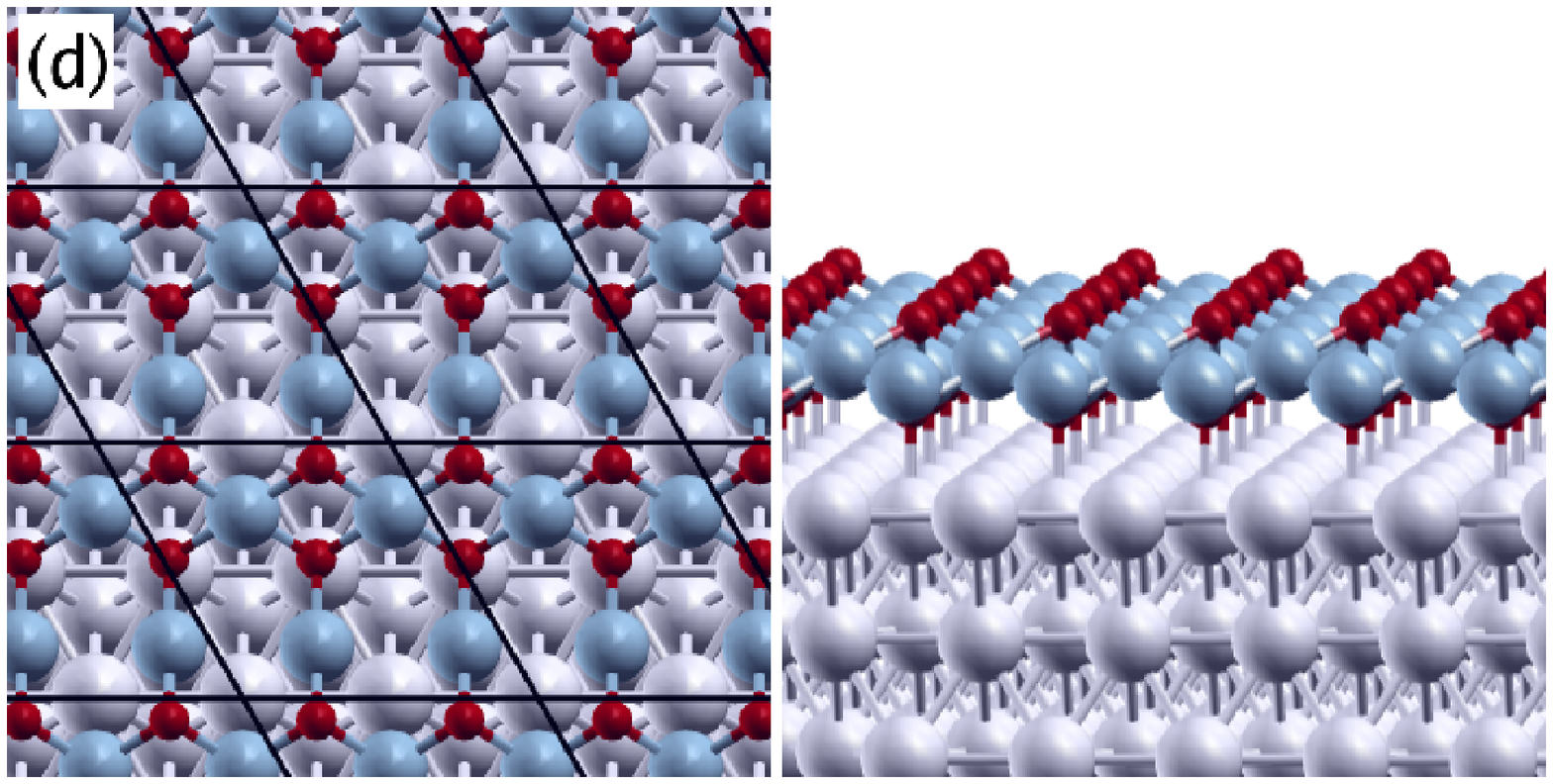}
\end{minipage}
\begin{minipage}[c]{0.45\textwidth}
\centering
\includegraphics[width=70mm, angle=0]{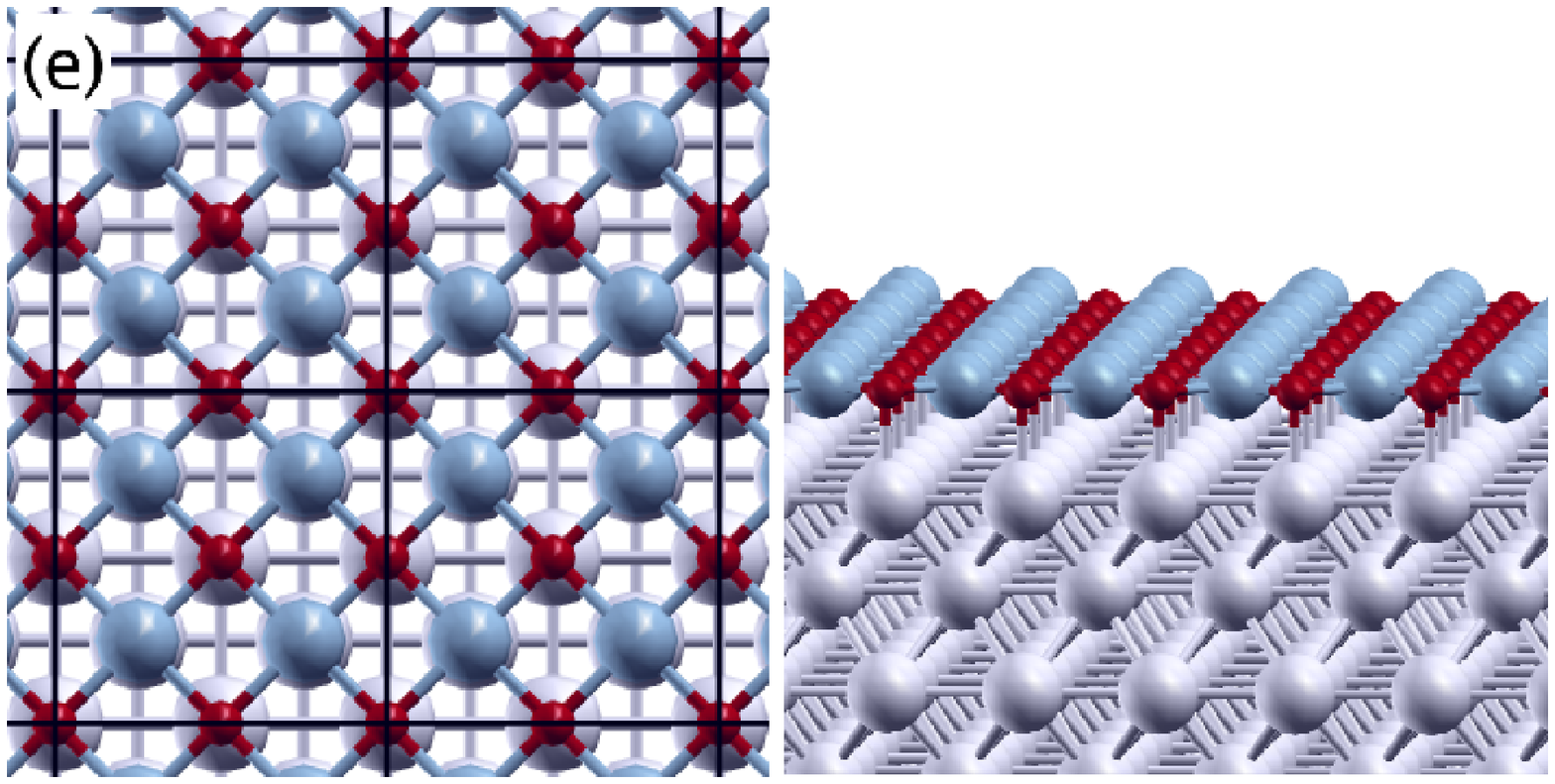}
\end{minipage}
\begin{minipage}[c]{0.45\textwidth}
\centering
\includegraphics[width=70mm, angle=0]{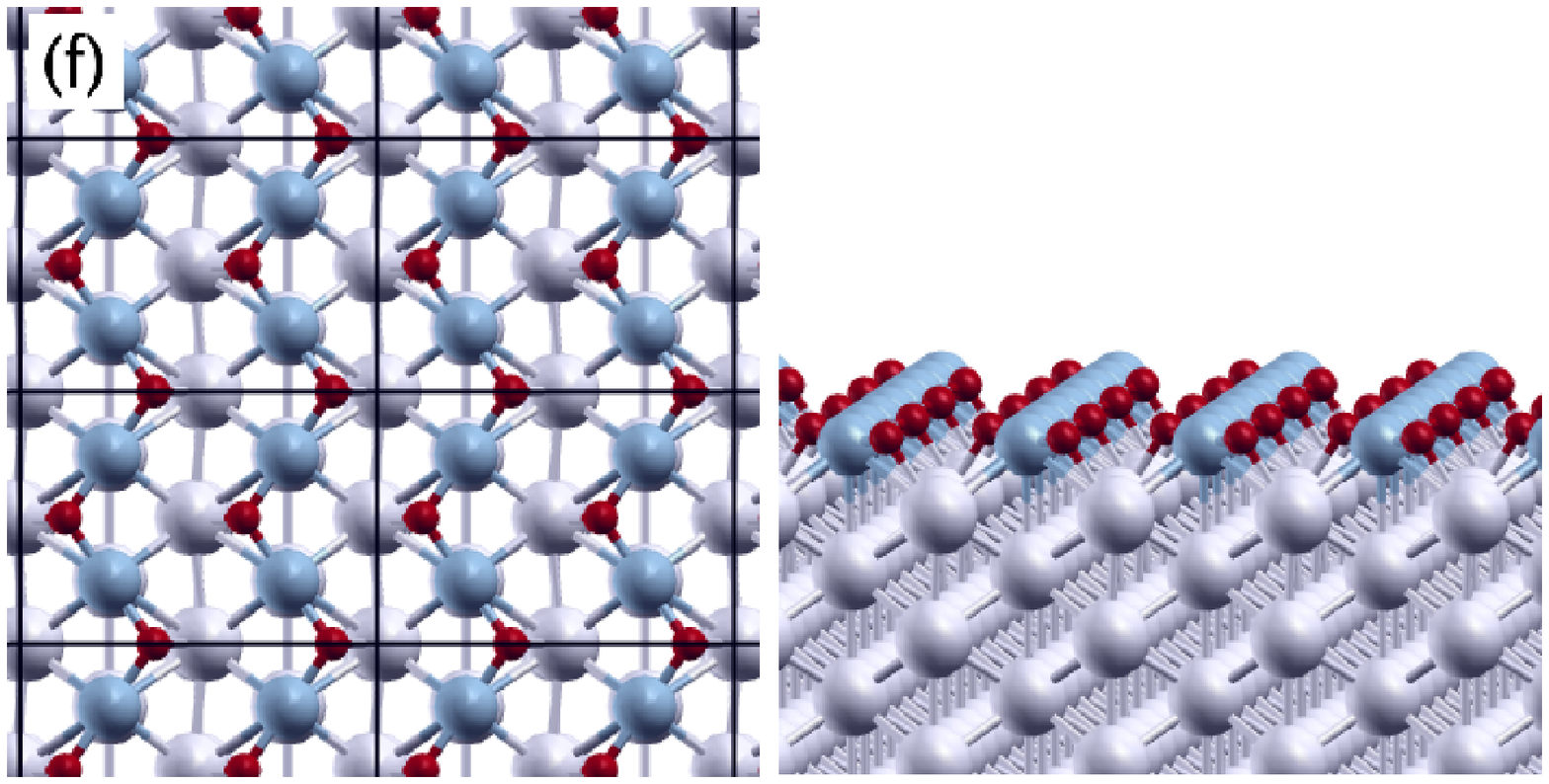}
\end{minipage}
\caption{\label{FigStruct} (Color online) Top and side view of some of the low energy structures considered in this work: (a) (2$\times$1) added row structure on the (110) surface, (b) ``$p$2'' on the (111) surface, (c) Ag/O(2$\times$2) on the (100) surface, (d) 1 layer of CuO on the (111) surface, (e) 1 layer of CuO on the (100) surface, (f) 1 layer of CuO on the (110) surface. The red (black) atoms represent oxygen, the blue (dark grey) ones represent copper and the grey (light grey) represent silver.}
\end{figure}  

For the first set of structures we employ (2$\times$2) unit cells and label the structures using the notation O$x$ML/Cu$y$ML, where $x$ and $y$ are the content of O and Cu, expressed in monolayers with respect to the underlying Ag surface. In the case of the (100) and (110) orientations, we also study the missing row and the added row reconstructions, respectively, allowing for Cu alloying in the first layer (see Fig.~\ref{FigStruct}(a)). For the (100) surface, in the absence of Cu, also the c(4$\times$6) reconstruction has been considered, using the data for the energetics reported in Ref.~\onlinecite{SerianiJPCM2008}. 

For the second set of structures, the geometries are derived from bulk Cu$_2$O. We use various film thicknesses, from 1 up to 5 layers. For each surface orientation, forcing the Cu$_2$O film to be commensurate to the underlying Ag surface ,results in a 7\% compression of the Cu$_2$O film area compared to the bulk value. In the case of the (111) orientation, we use the label ``$p$2'' or ``$p$4'' depending on whether the periodicity of the structure is (2$\times$2) or (4$\times$4) with respect to the clean Ag surface. Of particular interest are the ``$p$2'' and the ``$p$4-OCu$_3$'' structures. The atomic geometry of the ``$p$2'' structure is shown in Fig.~\ref{FigStruct}(b). Here each Cu is linearly bonded with two O atoms and each O atom is bonded to three Cu atoms, in a ring-like structure similar to the ones proposed for thin oxide-like layers on Ag(111)\cite{MichaelidesJVCT05} and Cu(111)\cite{SoonPRB2006}. The ``$p$4-OCu$_3$'' is a $p$4 structure in which an OCu$_3$ unit has been removed. 
For the (100) orientation we use a ($\sqrt{2}\times\sqrt{2}$) unit cell and  for the (110) surface we use a (2$\times$2) cell. None of these structures on the (100) and (110) surfaces, as we will show later, are found to be present in the phase diagram.
 
For the third set of structures, we consider thin layers of CuO-like configurations, with a thickness up to 5 layers. 
In this case we must bear in mind that bulk CuO is poorly described with DFT-PBE: CuO is a strongly correlated antiferromagnetic semiconductor, with a monoclinic structure ($a$=4.65 \AA, $b$=3.41 \AA, $c$=5.11 \AA, $\beta$=99.5$^\circ$), where Cu is linearly bonded to 2 O atoms and O is bound to 2 Cu atoms.~\cite{KimJACS03}
DFT-PBE, on the other hand, predicts CuO to be a metal with an almost orthorhombic structure ($a$=4.34 \AA, $b$=4.01 \AA, $c$=5.22 \AA, $\beta$=92.2$^\circ$), in which each O atom is tetrahedrally coordinated to 4 Cu atoms and each Cu atom is bonded in a square planar geometry to 4 O atoms.
The predictions for thick films of CuO must therefore be regarded as qualitative. The computed formation energy per oxygen atom in bulk CuO is 1.23 eV (experimental value 1.63 eV~\cite{CRC}) and the Cu-O bond length is about 1.97 \AA~(experimental value 1.67 \AA).
Going beyond the DFT-PBE description is a major task. Calculations for CuO and
Cu$_2$O using GGA+U plus G$_0$W$_0$,~\cite{JiangTBP} as well as other
methodology,~\cite{HuReutPRL2007} are in progress.
Given the fact that CuO does not have a cubic structure, forcing the CuO thin layers to be commensurate with the underlying Ag substrate leads to distortions that depend on the orientation of the surface. In our study we consider (2$\times$2) cells for the (111) and (100) surfaces and (2$\times$1) cells for the (110). This leads to a compression of the Cu-O bonds that, for every orientation, is of the order of 3\%, and the relaxed geometries of the structures strongly depend on the surface orientation (see Fig.~\ref{FigStruct}(d-f)).
We find that one layer of CuO is particularly stable for every orientation. For this structure we calculate the formation energy per oxygen atom to be 1.16 eV for the (111) surface and 1.26 eV for both the (100) and (110) surfaces.  

In addition to the structures considered in this work, it is likely that other structures involving thin oxide-like layers with similar surface energies exist for this system. Recent works on O/Ag(111)\cite{MichaelidesJVCT05,SchnadtPRL2006, SchmidPRL2006} and O/Pd(111)~\cite{KlikovitsPRB2007} have shown that for these systems a multitude of structures with similar energetics can be found. 

\subsection{Thermodynamic diagrams for the Cu-O/Ag system}

\begin{figure}[tb]
\begin{minipage}[c]{0.49\textwidth}
\centering
\includegraphics[width=62mm, angle=270]{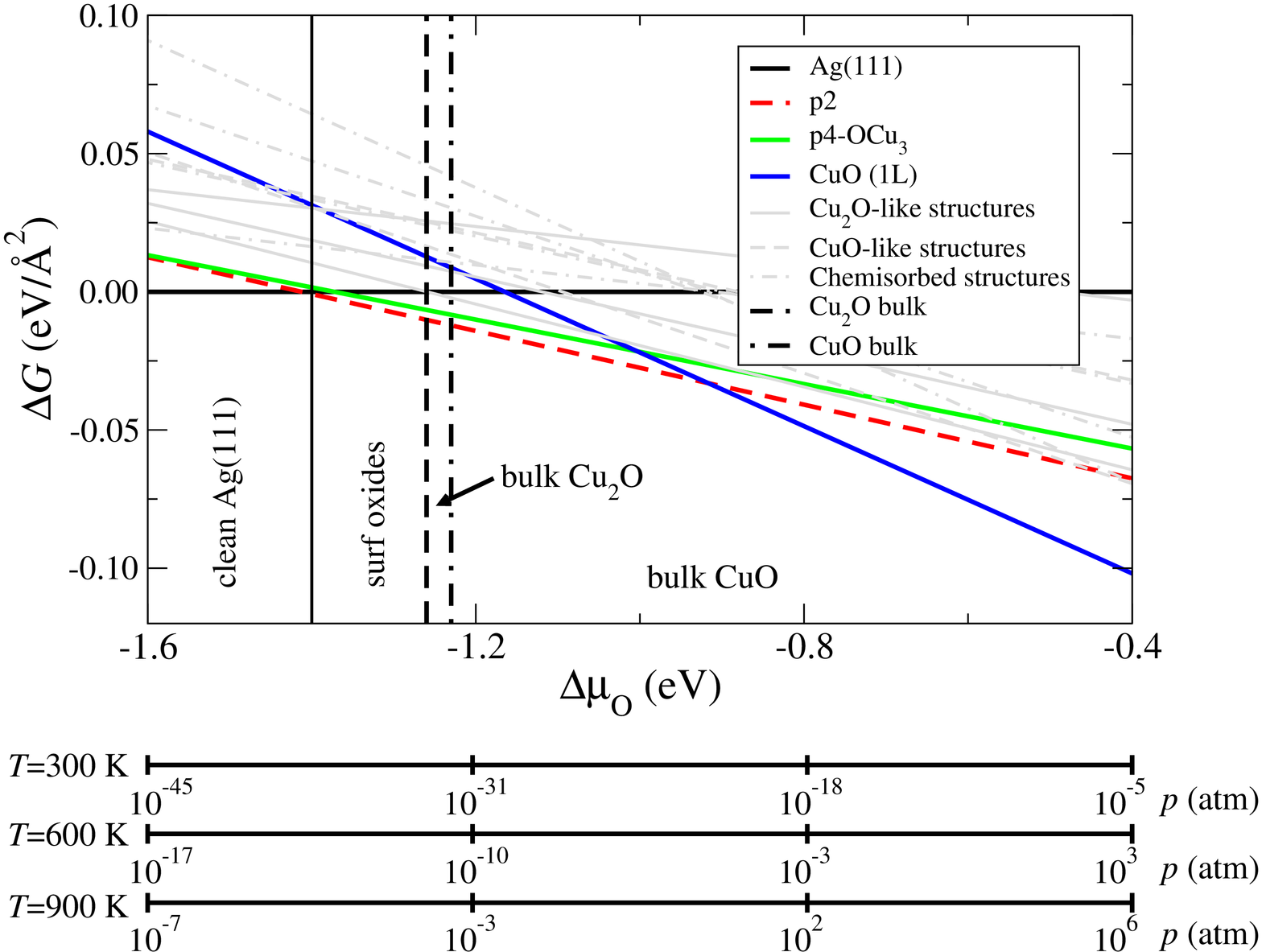}
\end{minipage}
\begin{minipage}[c]{0.49\textwidth}
\centering
\includegraphics[width=62mm, angle=270]{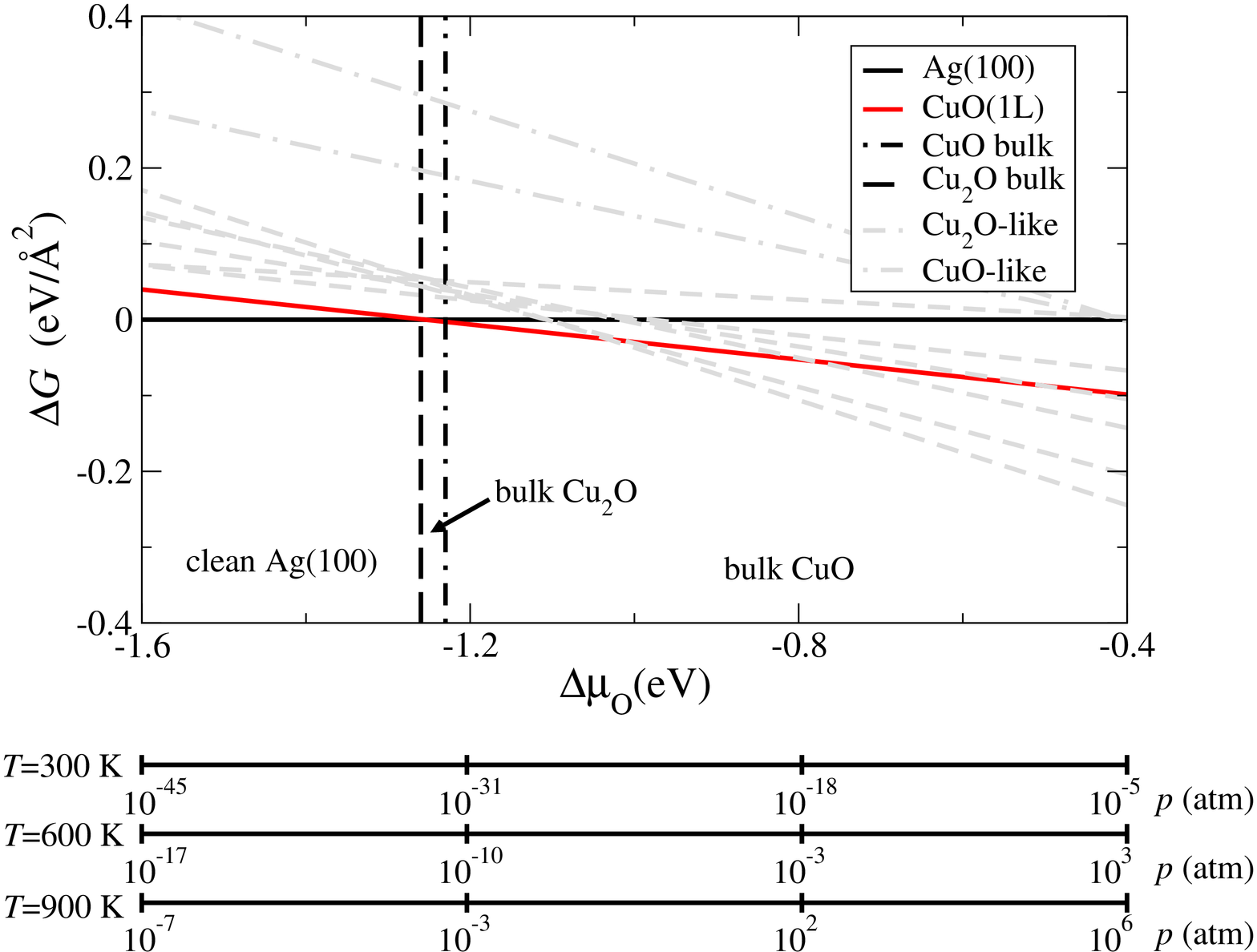}
\end{minipage}
\begin{minipage}[c]{1.0\textwidth}
\centering
\includegraphics[width=62mm, angle=270]{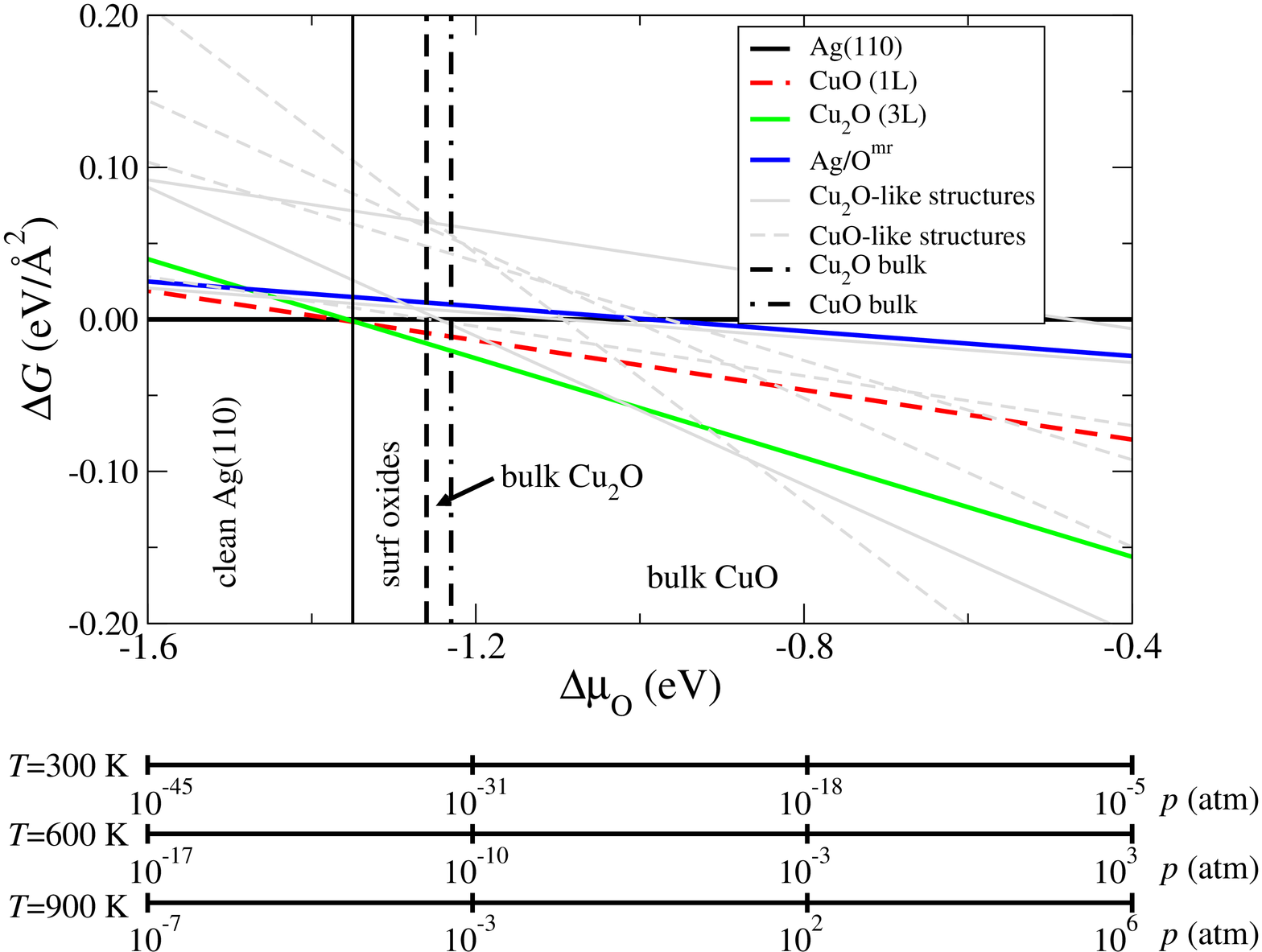}
\end{minipage}
\caption{\label{FigTD} Thermodynamic diagrams (i.e. change in surface free energy versus oxygen chemical potential) for the Ag/Cu/O system. The top left panel correspond to the (111) surface, the top right one to the (100) surface and the bottom one to the (110) surface. The vertical lines separate the regions where different structures are stable.}
\end{figure}

In Fig.~\ref{FigTD} we report the change in Gibbs surface free energy of the structures considered as a function of the change in oxygen chemical potential. The three graphs correspond to the three surface orientations considered in this work. We also show at the bottom of the plot, for three values of temperature (300, 600 and 900 K), the value of pressure corresponding to the chemical potential shown in the abscissa of the plot.

As we can see from Eq.~(\ref{eq_gamma3}), the slope of the lines in Fig.~\ref{FigTD} is proportional to the oxygen coverage, i.e. the higher the oxygen content the steeper the line. In all the three graphs, the vertical dashed line at $\Delta \mu_{\rm O} = -1.26$ eV represents the change in chemical potential above which Cu oxidizes to Cu$_2$O, which, as one can show, corresponds to the computed heat of formation of Cu$_2$O. The vertical dashed line at $\Delta \mu_{\rm O} = -1.23$ eV, on the other hand, represents the change in chemical potential above which Cu oxidizes to CuO. 

In the case of the (111) surface, we can see that in the range of values of the oxygen chemical potential considered ($-1.6 < \Delta\mu_{\rm O} < -0.4$ eV) the thermodynamically most stable structures are: Pure Ag for $\Delta\mu_{\rm O} < -1.43$ eV, the surface oxides ``$p$2'' and ``$p$4-OCu$_3$'' (almost degenerate) between  $-1.43 <\Delta \mu_{\rm O} < -1.26$ eV, bulk copper(I) oxide Cu$_2$O for $-1.26 <\Delta \mu_{\rm O} < -1.23$ eV and bulk copper(II) oxide CuO for $\Delta \mu_{\rm O} > -1.23$ eV. 

For the (100) surface we find that no surface oxides are stable. For $\Delta \mu_{\rm O} < -1.26$ eV pure Ag is the stable structure, while above that value of oxygen chemical potential, as in the (111) surface, we find the two bulk copper oxides to be stable. 

For the (110) surface, on the other hand, we find that in the range $-1.46 <\Delta \mu_{\rm O} < -1.26$ eV a three layer thick Cu$_2$O surface oxide is the stable structure, while at lower oxygen chemical potential the stable structure is pure Ag and at higher oxygen chemical potential the bulk copper oxides become stable.

It is interesting to note that none of the alloyed chemisorbed structures (the structures labelled as O$x$ML/Cu$y$ML) are thermodynamically stable for any of the three surfaces, while there are regions in which two-dimensional surface oxides are stable, both on the (111) and the (110) surfaces. This situation is similar to what has been found for the O/Cu(111)system,~\cite{SoonPRB2006} where it was argued that copper oxidation does not proceed via ordered chemisorbed structures, at variance with other transition metal surfaces. We also find that none of the O/Ag structures, i.e. structures containing no Cu (not shown in Fig.~\ref{FigTD}), are thermodynamically stable in the range considered here. This is not unexpected, given the weaker strength of the Ag-O bond with respect to the Cu-O one.

\subsection{Surface phase diagrams}

Having identified, among those considered here, the thermodynamically most stable ordered structures for the infinite (111), (100) and (110) surfaces, we now consider the situation in which we have a well defined Cu surface content in each of those surfaces. For a finite system (e.g. an Ag particle) of known dimensions and well defined Cu content, if we assume that all the Cu present in a spherical nanoparticle segregates and homogeneously distributes on the surfaces of the particle in a site of the Ag fcc lattice, given the size of particle and the total Cu content, we can estimate the Cu surface content. For example in a 50 nm spherical particle with a 1~\% Cu content (typical values of particle size and copper content of Ag-Cu alloys used in experiments in which the effect of Cu impurities was investigated~\cite{JankJC2005}) the (average) surface Cu content is 0.35 ML, where 1 ML corresponds to the first layer of the particle being pure copper. Experiments carried out on Ag-Cu alloys with 0.1-1$\%$ Cu content at $\sim$ 500 K have shown that in an oxidizing environment the Cu surface content is in the range 0.1-0.75 ML.~\cite{JankJC2005}

In the following, we will consider the three surfaces investigated in this work and use the results shown in Fig.~\ref{FigTD} to predict which structures will be present on the surfaces as a function of the copper content and the oxygen chemical potential. We also exploit the vast literature available for the O/Ag system~\cite{MichaelidesJVCT05, SchnadtPRL2006} to include the most stable structures for the system in the absence of copper.

\begin{figure}[tb]
\centering
\includegraphics[width=90mm, angle=270]{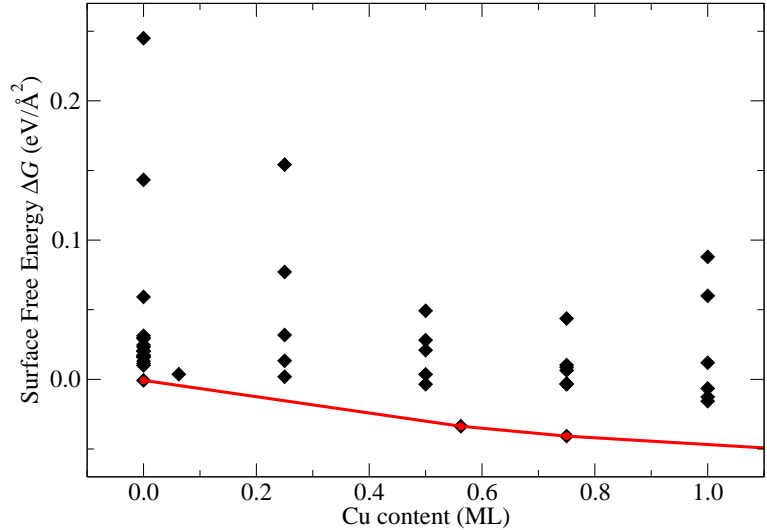}
\caption{\label{FigAg-Hull-0.8} (color online) Construction of the convex hull (indicated by the line) for the case of $\Delta\mu_{\rm O} = -0.80$ eV. Only the portion up to 1 ML is shown in the figure. The energy zero corresponds to the clean Ag(111) surface.}
\end{figure}

As an example, we show explicitly the case for $\Delta\mu_{\rm O} = -0.80$ eV, corresponding to an oxygen pressure of $10^{-3}$~atm. at the temperature of 600 K. Figure~\ref{FigAg-Hull-0.8} shows the change in surface free energy at $\Delta\mu_{\rm O} = -0.80$ eV of all the structures we have considered on the (111) surface (a total of 55 structures, including the O/Ag ones) as a function of the Cu surface content. By constructing the convex hull, i.e. the curve obtained by joining those structures that are most stable with respect to linear combinations of structures at other compositions that would yield the same total composition, we identify the structures (indicated by the red dots in Fig.~\ref{FigAg-Hull-0.8}) that are stable against phase separation into any two other structures. Although in Fig.~\ref{FigAg-Hull-0.8} we show only the portion of the diagram up to a Cu content of 1 ML, we have computed structures including up to five CuO and Cu$_2$O layers in order to be able to extrapolate the behavior of bulk CuO.

\begin{figure}[t]
\begin{minipage}[c]{0.45\textwidth}
\centering
\includegraphics[width=65mm, angle=270]{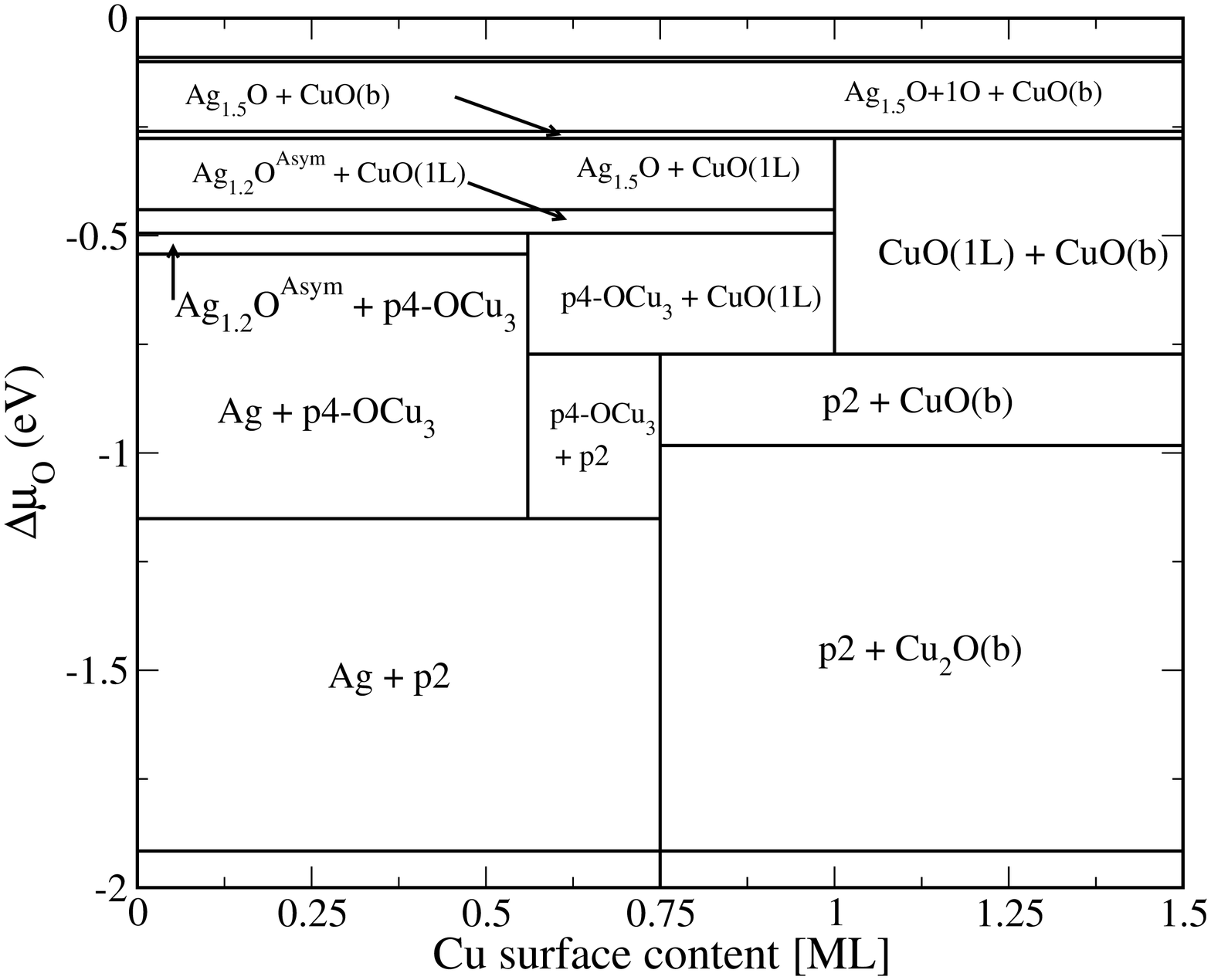}
\end{minipage}
\begin{minipage}[c]{0.45\textwidth}
\centering
\includegraphics[width=65mm, angle=270]{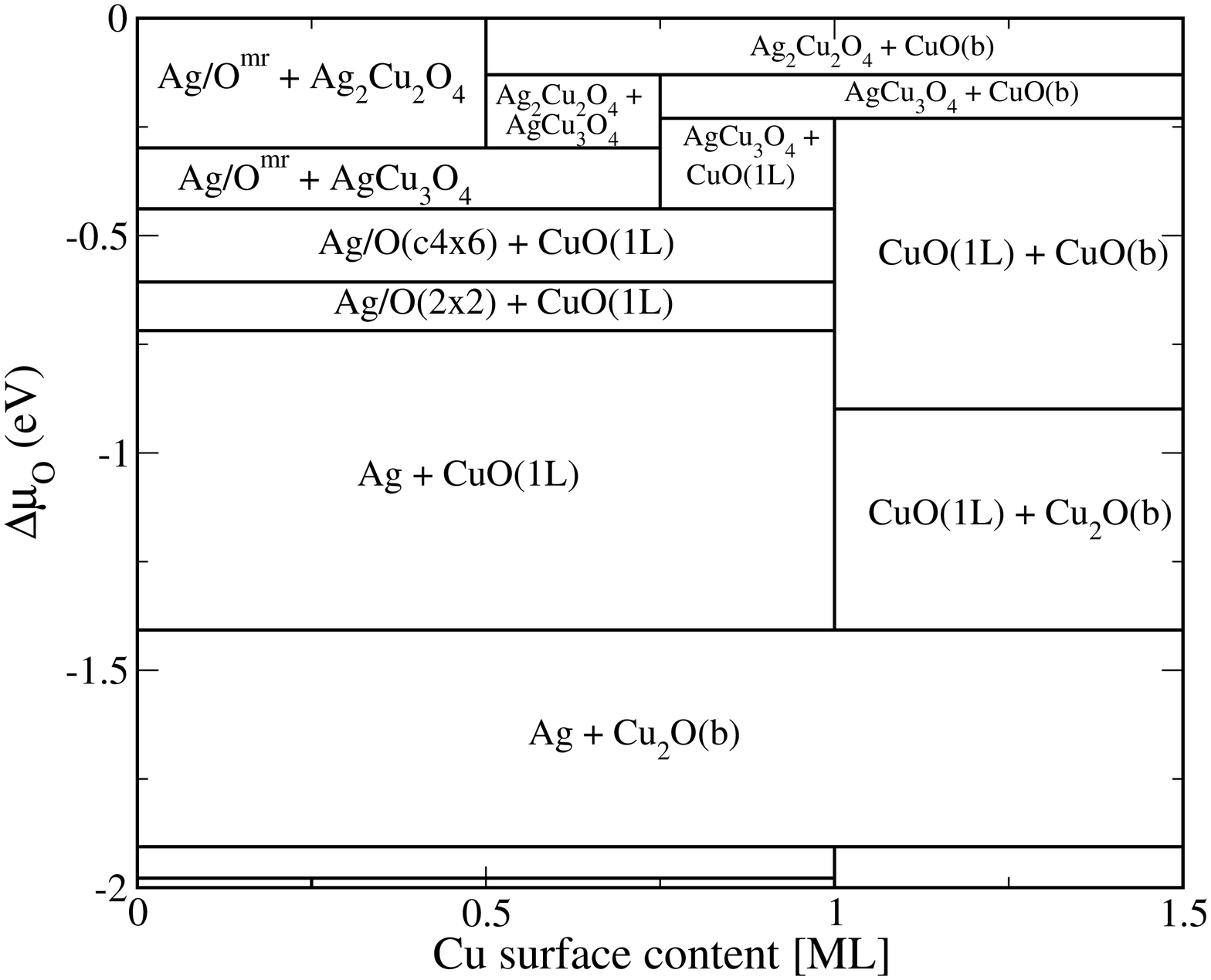}
\end{minipage}
\begin{minipage}[c]{0.45\textwidth}
\centering
\includegraphics[width=65mm, angle=270]{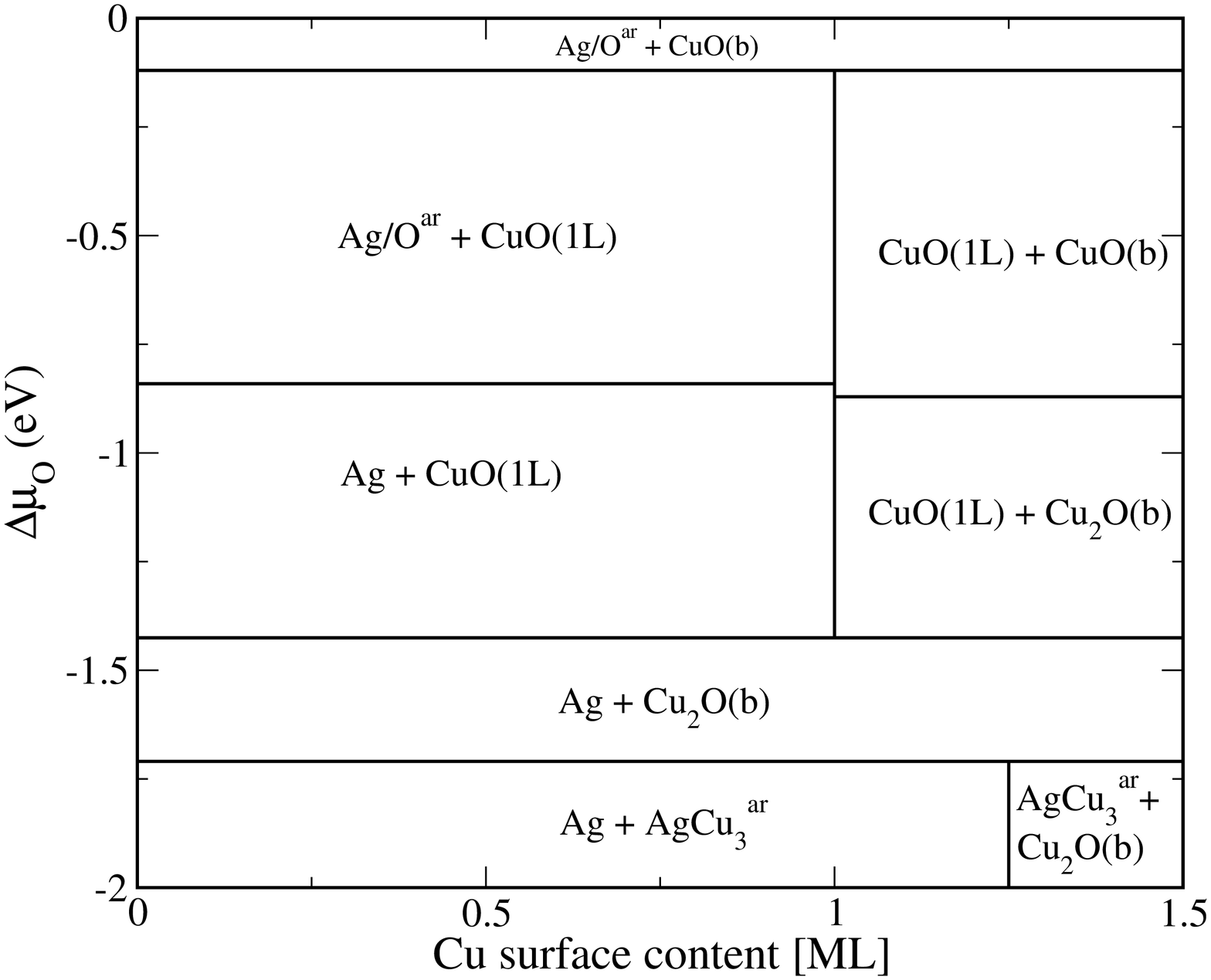}
\end{minipage}
\caption{\label{FigPD} Surface phase diagrams showing the structures belonging to the convex hull as a function of the Cu surface content and thechange in oxygen chemical potential. The top left panel refers to the (111) surface, the top right to the (100) surface and the bottom one to the (110) surface.}
\end{figure}

For a specific Cu content, e.g. 0.25 ML, the convex hull plot helps to predict that a mixture of pure Ag and ``$p$4-Cu$_3$O'' (rather than a single ordered structure with 0.25 ML Cu content) will be present on the Ag(111) surface at $\Delta\mu_{\rm O} = -0.80$~eV, in a ratio given by the lever rule, i.e. by the mass conservation law. By repeating this scheme for all the values of the oxygen chemical potential in the range considered in this work, we build a phase diagram as a function of the oxygen chemical potential and the copper surface content. This is shown in Fig.~\ref{FigPD} for the three surface orientations considered in this work. We find that the convex hull, for high Cu content, always includes the structure with the largest number of CuO layers (or Cu$_2$O layers at lower oxygen chemical potential), i.e. bulk oxide formation is favored beyond a certain Cu surface content. We therefore cut the plot in Fig.~\ref{FigPD} at a Cu composition of 1.50 ML, where it is understood that the rightmost structure is ``bulk'' oxide. However, care must be taken in the meaning given to such ``bulk'' structure in the context of a finite system as the one we are modelling here. We interpret the numerical evidence of the presence of the bulk structure in the convex hull as a tendency for the formation of thick patches of either CuO or Cu$_2$O.

In the phase diagram shown in Fig.~\ref{FigPD} we have explicitly written the two structures coexisting in a number of regions around the values of interest of oxygen chemical potential and copper content. Around $\Delta\mu_{\rm O} = -0.60$~eV, for all three surface orientations, we find that the stable structures include clean Ag (on the (111) surface) or Ag-O structures (Ag/O(c4$\times$6) on the (100) surface and Ag/O$^{\rm ar}$ on the (110) surface) at low Cu content, CuO(1L) for intermediate levels and bulk CuO at high Cu content. The fact that these features are general and do not depend strongly on the surface orientation allows us to draw some general conclusions. First of all, a 2D-alloy on the catalyst surface is never stable, for any of the low index surfaces examined here. The strength of the Cu-O bond is the factor determining the absence of any surface alloy and the presence, on the other hand, of thin or thick layers of copper oxide. Secondly, a thin layer of CuO is particularly favourable for all orientations, making this structure a possible candidate to explain the high selectivity of the Ag-Cu alloy for ethylene epoxidation.   

The picture emerging from our results is consistent with recent experiments performed on the Ag-Cu system under catalytic conditions.~\cite{SpirosUNP} In these experiments Ag-Cu nanopowders ($\sim$ 100 nm in diameter and 2.5\% Cu) were used as the catalyst for ethylene epoxidation at $T$ = 520 K and $p$ = 0.5 mbar, corresponding to $\Delta \mu_{\rm O} = -0.69$ eV. Through a combination of {\it in-situ} XPS, and Near Edge X-ray Absorption Fine Structure (NEXAFS) measurements, thin layers of CuO and Cu$_2$O are found be to present on the surface, while no signs on a surface alloy are detected. Areas of clean Ag are also present on the surface, in agreement with our theoretical calculations.

These results also show that the simple 2D-alloy structure adopted in Ref.~\onlinecite{LinicJC2004} for the Ag-Cu surface to model theoretically the ethylene epoxidation reaction is not stable and is significantly different from what we expect to be relevant under high pressure conditions. The model used in Ref.~\onlinecite{LinicJC2004} assumes a (2$\times$2) periodicity for the Ag(111) surface in which one atom out of four is replaced with copper, and oxygen is chemisorbed on it. Our results suggest on the other hand that, for this surface, a model that includes, depending on the Cu surface content, clean Ag(111), patches of oxide-like thin layers and thick layers of CuO, is a more appropriate model for the Ag-Cu catalyst (111) surface. Moreover, we have shown that this picture is valid also for the other low index surfaces.
\section{Conclusions}
\label{Concl}
By means of {\it ab-initio} atomistic thermodynamics we have investigated the low index surfaces of the Ag-Cu alloy in equilibrium with an oxygen atmosphere, accounting for the effects of temperature and pressure. For all the surface orientations considered in this work, the presence of oxygen on the surface leads to copper segregation, due to the strength of the O-Cu bond relative to the O-Ag one. We have studied, for every orientation, surface structures that include chemisorbed oxygen on a 2D Ag-Cu alloy as well as Cu$_2$O and CuO layers, both in the single layer limit and in the bulk limit. Through the construction of the oxygen chemical potential dependent convex hull, we have predicted, as a function of the surface copper content, temperature and pressure, the combinations of structures that are most stable. The phase diagrams constructed this way show remarkable similarities for different surface orientations. For the (111) surface, around the temperature and pressure of interest for practical applications and depending on the copper surface content, our results indicate that clean Ag(111) and thin copper oxide-like structures ($p$4-Cu$_3$O, $p$2 and CuO(1L)) can coexist. For the (100) orientation the Ag/O(2$\times$2) and the CuO(1L) structures are predicted to be present on the surface, while on the (110) surface the Ag/O$^{\rm mr}$ and the CuO(1L) structures are present in the phase diagram around the region of interest. Therefore a combination of Ag/O structures and thin copper oxide layers are predicted for all surface orientations. A single layer of CuO, in particular, has a wide range of stability on all surfaces. The model for the surfaces of the Ag-Cu catalyst that emerges from our calculations differs substantially from the one proposed in earlier works aimed at investigating the effect of copper impurities in silver on the mechanism of ethylene epoxidation. Our results suggest that to gain insight into the full catalytic cycle one should consider oxygen species such as those in the proposed structures, rather than just oxygen chemisorbed on the clean alloy surface.

Here we have limited ourselves to the study of the thermodynamic equilibrium between the Ag-Cu alloy and an oxygen atmosphere. In the steady state conditions of catalysis, however, the surface of the alloy could be altered by the interplay of various processes, such as molecular adsorption and desorption and chemical reactions happening on the surface. Furthermore, in our thermodynamic equilibrium we only accounted for the presence of oxygen, while neglecting the reducing effects of ethylene. Our work can therefore be considered only as a first, albeit necessary, step toward understanding the full catalytic cycle.

\bibliography{main}

\begin{thebibliography}{59}
\expandafter\ifx\csname natexlab\endcsname\relax\def\natexlab#1{#1}\fi
\expandafter\ifx\csname bibnamefont\endcsname\relax
  \def\bibnamefont#1{#1}\fi
\expandafter\ifx\csname bibfnamefont\endcsname\relax
  \def\bibfnamefont#1{#1}\fi
\expandafter\ifx\csname citenamefont\endcsname\relax
  \def\citenamefont#1{#1}\fi
\expandafter\ifx\csname url\endcsname\relax
  \def\url#1{\texttt{#1}}\fi
\expandafter\ifx\csname urlprefix\endcsname\relax\def\urlprefix{URL }\fi
\providecommand{\bibinfo}[2]{#2}
\providecommand{\eprint}[2][]{\url{#2}}

\bibitem[{\citenamefont{Stampfl et~al.}(2002)\citenamefont{Stampfl,
  Ganduglia-Pirovano, Reuter, and Scheffler}}]{StampflSS2002}
\bibinfo{author}{\bibfnamefont{C.}~\bibnamefont{Stampfl}},
  \bibinfo{author}{\bibfnamefont{M.~V.} \bibnamefont{Ganduglia-Pirovano}},
  \bibinfo{author}{\bibfnamefont{K.}~\bibnamefont{Reuter}}, \bibnamefont{and}
  \bibinfo{author}{\bibfnamefont{M.}~\bibnamefont{Scheffler}},
  \bibinfo{journal}{Surf. Sci.} \textbf{\bibinfo{volume}{500}},
  \bibinfo{pages}{368} (\bibinfo{year}{2002}).

\bibitem[{\citenamefont{Reuter et~al.}(2005)\citenamefont{Reuter, Stampfl, and
  Scheffler}}]{HandbookMM}
\bibinfo{author}{\bibfnamefont{K.}~\bibnamefont{Reuter}},
  \bibinfo{author}{\bibfnamefont{C.}~\bibnamefont{Stampfl}}, \bibnamefont{and}
  \bibinfo{author}{\bibfnamefont{M.}~\bibnamefont{Scheffler}},
  \emph{\bibinfo{title}{Handbook of Materials Modeling, Part A. Methods}},
  \bibinfo{number}{p. 149-234} (\bibinfo{publisher}{Springer, Berlin},
  \bibinfo{year}{2005}).

\bibitem[{\citenamefont{Hohenberg and Kohn}(1964)}]{HK64}
\bibinfo{author}{\bibfnamefont{P.}~\bibnamefont{Hohenberg}} \bibnamefont{and}
  \bibinfo{author}{\bibfnamefont{W.}~\bibnamefont{Kohn}},
  \bibinfo{journal}{Phys. Rev.} \textbf{\bibinfo{volume}{136}},
  \bibinfo{pages}{B864} (\bibinfo{year}{1964}).

\bibitem[{\citenamefont{Kohn and Sham}(1965)}]{KS65}
\bibinfo{author}{\bibfnamefont{W.}~\bibnamefont{Kohn}} \bibnamefont{and}
  \bibinfo{author}{\bibfnamefont{L.}~\bibnamefont{Sham}},
  \bibinfo{journal}{Phys. Rev.} \textbf{\bibinfo{volume}{140}},
  \bibinfo{pages}{A1133} (\bibinfo{year}{1965}).

\bibitem[{\citenamefont{Stampfl}(2005)}]{StampflCT2005}
\bibinfo{author}{\bibfnamefont{C.}~\bibnamefont{Stampfl}},
  \bibinfo{journal}{Catal. Today} \textbf{\bibinfo{volume}{105}},
  \bibinfo{pages}{17} (\bibinfo{year}{2005}).

\bibitem[{\citenamefont{Ertl}(2008)}]{ErtlACIE2008}
\bibinfo{author}{\bibfnamefont{G.}~\bibnamefont{Ertl}},
  \bibinfo{journal}{Angew. Chem. Int. Ed.} \textbf{\bibinfo{volume}{47}},
  \bibinfo{pages}{3524} (\bibinfo{year}{2008}).

\bibitem[{\citenamefont{Reuter et~al.}(2002)\citenamefont{Reuter, Stampfl,
  Ganduglia-Pirovano, and Scheffler}}]{ReuterCPL02}
\bibinfo{author}{\bibfnamefont{K.}~\bibnamefont{Reuter}},
  \bibinfo{author}{\bibfnamefont{C.}~\bibnamefont{Stampfl}},
  \bibinfo{author}{\bibfnamefont{M.~V.} \bibnamefont{Ganduglia-Pirovano}},
  \bibnamefont{and}
  \bibinfo{author}{\bibfnamefont{M.}~\bibnamefont{Scheffler}},
  \bibinfo{journal}{Chem. Phys. Lett.} \textbf{\bibinfo{volume}{352}},
  \bibinfo{pages}{311} (\bibinfo{year}{2002}).

\bibitem[{\citenamefont{Hendriksen et~al.}(2005)\citenamefont{Hendriksen,
  Bobaru, and Frenkel}}]{FrenkelCAT05}
\bibinfo{author}{\bibfnamefont{B.~L.~M.} \bibnamefont{Hendriksen}},
  \bibinfo{author}{\bibfnamefont{S.~C.} \bibnamefont{Bobaru}},
  \bibnamefont{and} \bibinfo{author}{\bibfnamefont{J.~W.~M.}
  \bibnamefont{Frenkel}}, \bibinfo{journal}{Topics in Catalysis}
  \textbf{\bibinfo{volume}{36}}, \bibinfo{pages}{43} (\bibinfo{year}{2005}).

\bibitem[{\citenamefont{Reuter and Scheffler}(2003)}]{ReuterPRL03}
\bibinfo{author}{\bibfnamefont{K.}~\bibnamefont{Reuter}} \bibnamefont{and}
  \bibinfo{author}{\bibfnamefont{M.}~\bibnamefont{Scheffler}},
  \bibinfo{journal}{Phys. Rev. Lett.} \textbf{\bibinfo{volume}{90}},
  \bibinfo{pages}{046103} (\bibinfo{year}{2003}).

\bibitem[{\citenamefont{Reuter and Scheffler}(2004)}]{ReuterPRL04}
\bibinfo{author}{\bibfnamefont{K.}~\bibnamefont{Reuter}} \bibnamefont{and}
  \bibinfo{author}{\bibfnamefont{M.}~\bibnamefont{Scheffler}},
  \bibinfo{journal}{Phys. Rev. Lett.} \textbf{\bibinfo{volume}{93}},
  \bibinfo{pages}{116105} (\bibinfo{year}{2004}).

\bibitem[{\citenamefont{Rogal et~al.}(2007)\citenamefont{Rogal, Reuter, and
  Scheffler}}]{RogalPRL07}
\bibinfo{author}{\bibfnamefont{J.}~\bibnamefont{Rogal}},
  \bibinfo{author}{\bibfnamefont{K.}~\bibnamefont{Reuter}}, \bibnamefont{and}
  \bibinfo{author}{\bibfnamefont{M.}~\bibnamefont{Scheffler}},
  \bibinfo{journal}{Phys. Rev. Lett.} \textbf{\bibinfo{volume}{98}},
  \bibinfo{pages}{046101} (\bibinfo{year}{2007}).

\bibitem[{\citenamefont{Sinfelt}(1983)}]{SinfeltBOOK}
\bibinfo{author}{\bibfnamefont{J.~H.} \bibnamefont{Sinfelt}},
  \emph{\bibinfo{title}{Bimetallic Catalysts: Discoveries, Concepts and
  Application}} (\bibinfo{publisher}{Wiley, New York}, \bibinfo{year}{1983}).

\bibitem[{\citenamefont{Liu and N\o{}rskov}(2001)}]{Liu01}
\bibinfo{author}{\bibfnamefont{P.}~\bibnamefont{Liu}} \bibnamefont{and}
  \bibinfo{author}{\bibfnamefont{J.~K.} \bibnamefont{N\o{}rskov}},
  \bibinfo{journal}{Phys. Chem. Chem. Phys.} \textbf{\bibinfo{volume}{3}},
  \bibinfo{pages}{3814} (\bibinfo{year}{2001}).

\bibitem[{\citenamefont{Greeley and Mavrikakis}(2004)}]{Greeley04}
\bibinfo{author}{\bibfnamefont{J.}~\bibnamefont{Greeley}} \bibnamefont{and}
  \bibinfo{author}{\bibfnamefont{M.}~\bibnamefont{Mavrikakis}},
  \bibinfo{journal}{Nature Mat.} \textbf{\bibinfo{volume}{3}},
  \bibinfo{pages}{810} (\bibinfo{year}{2004}).

\bibitem[{\citenamefont{Greeley et~al.}(2006)\citenamefont{Greeley, Jaramillo,
  Bonde, Chorkendorff, and N\o{}rskov}}]{Greeley06}
\bibinfo{author}{\bibfnamefont{J.}~\bibnamefont{Greeley}},
  \bibinfo{author}{\bibfnamefont{T.~F.} \bibnamefont{Jaramillo}},
  \bibinfo{author}{\bibfnamefont{J.}~\bibnamefont{Bonde}},
  \bibinfo{author}{\bibfnamefont{I.}~\bibnamefont{Chorkendorff}},
  \bibnamefont{and}
  \bibinfo{author}{\bibfnamefont{J.}~\bibnamefont{N\o{}rskov}},
  \bibinfo{journal}{Nature Mat.} \textbf{\bibinfo{volume}{5}},
  \bibinfo{pages}{909} (\bibinfo{year}{2006}).

\bibitem[{\citenamefont{van Santen and Kuipers}(1987)}]{Santen87}
\bibinfo{author}{\bibfnamefont{R.~A.} \bibnamefont{van Santen}}
  \bibnamefont{and} \bibinfo{author}{\bibfnamefont{H.~P. C.~E.}
  \bibnamefont{Kuipers}}, \bibinfo{journal}{Adv. Catal.}
  \textbf{\bibinfo{volume}{35}}, \bibinfo{pages}{265} (\bibinfo{year}{1987}).

\bibitem[{\citenamefont{Linic and Barteau}(2002)}]{LinicJACS2002}
\bibinfo{author}{\bibfnamefont{S.}~\bibnamefont{Linic}} \bibnamefont{and}
  \bibinfo{author}{\bibfnamefont{M.~A.} \bibnamefont{Barteau}},
  \bibinfo{journal}{J. Am. Chem. Soc.} \textbf{\bibinfo{volume}{124}},
  \bibinfo{pages}{310} (\bibinfo{year}{2002}).

\bibitem[{\citenamefont{Linic and Barteau}(2003)}]{LinicJACS2003}
\bibinfo{author}{\bibfnamefont{S.}~\bibnamefont{Linic}} \bibnamefont{and}
  \bibinfo{author}{\bibfnamefont{M.~A.} \bibnamefont{Barteau}},
  \bibinfo{journal}{J. Am. Chem. Soc.} \textbf{\bibinfo{volume}{125}},
  \bibinfo{pages}{4034} (\bibinfo{year}{2003}).

\bibitem[{\citenamefont{Linic et~al.}(2004)\citenamefont{Linic, Jankowiak, and
  Barteau}}]{LinicJC2004}
\bibinfo{author}{\bibfnamefont{S.}~\bibnamefont{Linic}},
  \bibinfo{author}{\bibfnamefont{J.}~\bibnamefont{Jankowiak}},
  \bibnamefont{and} \bibinfo{author}{\bibfnamefont{M.~A.}
  \bibnamefont{Barteau}}, \bibinfo{journal}{J. Catal.}
  \textbf{\bibinfo{volume}{224}}, \bibinfo{pages}{489} (\bibinfo{year}{2004}).

\bibitem[{\citenamefont{Jankowiak and Barteau}(2005)}]{JankJC2005}
\bibinfo{author}{\bibfnamefont{J.~T.} \bibnamefont{Jankowiak}}
  \bibnamefont{and} \bibinfo{author}{\bibfnamefont{M.~A.}
  \bibnamefont{Barteau}}, \bibinfo{journal}{J. Catal.}
  \textbf{\bibinfo{volume}{236}}, \bibinfo{pages}{366} (\bibinfo{year}{2005}).

\bibitem[{\citenamefont{Torres et~al.}(2005)\citenamefont{Torres, Lopez, Illas,
  and Lambert}}]{TorresJACS2005}
\bibinfo{author}{\bibfnamefont{D.}~\bibnamefont{Torres}},
  \bibinfo{author}{\bibfnamefont{N.}~\bibnamefont{Lopez}},
  \bibinfo{author}{\bibfnamefont{F.}~\bibnamefont{Illas}}, \bibnamefont{and}
  \bibinfo{author}{\bibfnamefont{R.}~\bibnamefont{Lambert}},
  \bibinfo{journal}{J. Am. Chem. Soc.} \textbf{\bibinfo{volume}{127}},
  \bibinfo{pages}{10774} (\bibinfo{year}{2005}).

\bibitem[{\citenamefont{Kokalj et~al.}(2008)\citenamefont{Kokalj, Gava,
  de~Gironcoli, and Baroni}}]{KokaljJC2008}
\bibinfo{author}{\bibfnamefont{A.}~\bibnamefont{Kokalj}},
  \bibinfo{author}{\bibfnamefont{P.}~\bibnamefont{Gava}},
  \bibinfo{author}{\bibfnamefont{S.}~\bibnamefont{de~Gironcoli}},
  \bibnamefont{and} \bibinfo{author}{\bibfnamefont{S.}~\bibnamefont{Baroni}},
  \bibinfo{journal}{J. Catal} \textbf{\bibinfo{volume}{254}},
  \bibinfo{pages}{304} (\bibinfo{year}{2008}).

\bibitem[{\citenamefont{Michaelides et~al.}(2005)\citenamefont{Michaelides,
  Reuter, and Scheffler}}]{MichaelidesJVCT05}
\bibinfo{author}{\bibfnamefont{A.}~\bibnamefont{Michaelides}},
  \bibinfo{author}{\bibfnamefont{K.}~\bibnamefont{Reuter}}, \bibnamefont{and}
  \bibinfo{author}{\bibfnamefont{M.}~\bibnamefont{Scheffler}},
  \bibinfo{journal}{J. Vac. Sci. Technol. A} \textbf{\bibinfo{volume}{23}},
  \bibinfo{pages}{1487} (\bibinfo{year}{2005}).

\bibitem[{\citenamefont{Schnadt et~al.}(2006)\citenamefont{Schnadt,
  Michaelides, Knudsen, Vang, Reuter, Laegsgaard, Scheffler, and
  Besenbacher}}]{SchnadtPRL2006}
\bibinfo{author}{\bibfnamefont{J.}~\bibnamefont{Schnadt}},
  \bibinfo{author}{\bibfnamefont{A.}~\bibnamefont{Michaelides}},
  \bibinfo{author}{\bibfnamefont{J.}~\bibnamefont{Knudsen}},
  \bibinfo{author}{\bibfnamefont{R.~T.} \bibnamefont{Vang}},
  \bibinfo{author}{\bibfnamefont{K.}~\bibnamefont{Reuter}},
  \bibinfo{author}{\bibfnamefont{E.}~\bibnamefont{Laegsgaard}},
  \bibinfo{author}{\bibfnamefont{M.}~\bibnamefont{Scheffler}},
  \bibnamefont{and}
  \bibinfo{author}{\bibfnamefont{F.}~\bibnamefont{Besenbacher}},
  \bibinfo{journal}{Phys. Rev. Lett.} \textbf{\bibinfo{volume}{96}},
  \bibinfo{pages}{146101} (\bibinfo{year}{2006}).

\bibitem[{\citenamefont{Schmid et~al.}(2006)\citenamefont{Schmid, Reicho,
  Stierle, Costina, Klikovits, Kostelnik, Dubay, Kresse, Gustafson, Lundgren
  et~al.}}]{SchmidPRL2006}
\bibinfo{author}{\bibfnamefont{M.}~\bibnamefont{Schmid}},
  \bibinfo{author}{\bibfnamefont{A.}~\bibnamefont{Reicho}},
  \bibinfo{author}{\bibfnamefont{A.}~\bibnamefont{Stierle}},
  \bibinfo{author}{\bibfnamefont{I.}~\bibnamefont{Costina}},
  \bibinfo{author}{\bibfnamefont{J.}~\bibnamefont{Klikovits}},
  \bibinfo{author}{\bibfnamefont{P.}~\bibnamefont{Kostelnik}},
  \bibinfo{author}{\bibfnamefont{O.}~\bibnamefont{Dubay}},
  \bibinfo{author}{\bibfnamefont{G.}~\bibnamefont{Kresse}},
  \bibinfo{author}{\bibfnamefont{J.}~\bibnamefont{Gustafson}},
  \bibinfo{author}{\bibfnamefont{E.}~\bibnamefont{Lundgren}},
  \bibnamefont{et~al.}, \bibinfo{journal}{Phys. Rev. Lett.}
  \textbf{\bibinfo{volume}{96}}, \bibinfo{pages}{146102}
  (\bibinfo{year}{2006}).

\bibitem[{\citenamefont{Perdew et~al.}(1996)\citenamefont{Perdew, Burke, and
  Ernzerhof}}]{PBE}
\bibinfo{author}{\bibfnamefont{J.~P.} \bibnamefont{Perdew}},
  \bibinfo{author}{\bibfnamefont{K.}~\bibnamefont{Burke}}, \bibnamefont{and}
  \bibinfo{author}{\bibfnamefont{M.}~\bibnamefont{Ernzerhof}},
  \bibinfo{journal}{Phys. Rev. Lett.} \textbf{\bibinfo{volume}{77}},
  \bibinfo{pages}{3865} (\bibinfo{year}{1996}).

\bibitem[{\citenamefont{Vanderbilt}(1990)}]{USPP}
\bibinfo{author}{\bibfnamefont{D.}~\bibnamefont{Vanderbilt}},
  \bibinfo{journal}{Phys. Rev. B} \textbf{\bibinfo{volume}{41}},
  \bibinfo{pages}{7892} (\bibinfo{year}{1990}).

\bibitem[{USe()}]{USespresso}
\bibinfo{note}{The ultrasoft pseudopotentials for Ag, Cu and O were taken from
  the PWscf pseudopotential download page: http://www.pwscf.org/pseudo/htm
  (files: Ag.pbe-d-rrkjus.UPF, Cu.pbe-d-rrkjus.UPF, O.pbe-rrkjus.UPF)}.

\bibitem[{\citenamefont{Monkhorst and Pack}(1976)}]{Monkhorst-Pack}
\bibinfo{author}{\bibfnamefont{H.~J.} \bibnamefont{Monkhorst}}
  \bibnamefont{and} \bibinfo{author}{\bibfnamefont{J.~D.} \bibnamefont{Pack}},
  \bibinfo{journal}{Phys. Rev. B} \textbf{\bibinfo{volume}{13}},
  \bibinfo{pages}{5188} (\bibinfo{year}{1976}).

\bibitem[{\citenamefont{Marzari et~al.}(1999)\citenamefont{Marzari, Vanderbilt,
  De~Vita, and Payne}}]{MVcold}
\bibinfo{author}{\bibfnamefont{N.}~\bibnamefont{Marzari}},
  \bibinfo{author}{\bibfnamefont{D.}~\bibnamefont{Vanderbilt}},
  \bibinfo{author}{\bibfnamefont{A.}~\bibnamefont{De~Vita}}, \bibnamefont{and}
  \bibinfo{author}{\bibfnamefont{M.~C.} \bibnamefont{Payne}},
  \bibinfo{journal}{Phys. Rev. Lett} \textbf{\bibinfo{volume}{82}},
  \bibinfo{pages}{3296} (\bibinfo{year}{1999}).

\bibitem[{\citenamefont{Baroni et~al.}()\citenamefont{Baroni, Corso,
  de~Gironcoli, Giannozzi, Cavazzoni, Ballabio, Scandolo, Chiarotti, Focher,
  Pasquarello et~al.}}]{espresso}
\bibinfo{author}{\bibfnamefont{S.}~\bibnamefont{Baroni}},
  \bibinfo{author}{\bibfnamefont{A.~D.} \bibnamefont{Corso}},
  \bibinfo{author}{\bibfnamefont{S.}~\bibnamefont{de~Gironcoli}},
  \bibinfo{author}{\bibfnamefont{P.}~\bibnamefont{Giannozzi}},
  \bibinfo{author}{\bibfnamefont{C.}~\bibnamefont{Cavazzoni}},
  \bibinfo{author}{\bibfnamefont{G.}~\bibnamefont{Ballabio}},
  \bibinfo{author}{\bibfnamefont{S.}~\bibnamefont{Scandolo}},
  \bibinfo{author}{\bibfnamefont{G.}~\bibnamefont{Chiarotti}},
  \bibinfo{author}{\bibfnamefont{P.}~\bibnamefont{Focher}},
  \bibinfo{author}{\bibfnamefont{A.}~\bibnamefont{Pasquarello}},
  \bibnamefont{et~al.}, \emph{\bibinfo{title}{Quantum-espresso}},
  \bibinfo{note}{http://www.pwscf.org}.

\bibitem[{\citenamefont{Piccinin et~al.}(2008)\citenamefont{Piccinin, Stampfl,
  and Scheffler}}]{PiccininPRB2008}
\bibinfo{author}{\bibfnamefont{S.}~\bibnamefont{Piccinin}},
  \bibinfo{author}{\bibfnamefont{C.}~\bibnamefont{Stampfl}}, \bibnamefont{and}
  \bibinfo{author}{\bibfnamefont{M.}~\bibnamefont{Scheffler}},
  \bibinfo{journal}{Phys. Rev. B} \textbf{\bibinfo{volume}{77}},
  \bibinfo{pages}{075426} (\bibinfo{year}{2008}).

\bibitem[{\citenamefont{Huber and Herzberg}(1979)}]{HuberBOOK}
\bibinfo{author}{\bibfnamefont{K.~P.} \bibnamefont{Huber}} \bibnamefont{and}
  \bibinfo{author}{\bibfnamefont{G.}~\bibnamefont{Herzberg}},
  \emph{\bibinfo{title}{Molecular Spectra and Molecular Structure IV: Constants
  of Diatomic Molecules}} (\bibinfo{publisher}{Van Nostrand Reinhold, New
  York}, \bibinfo{year}{1979}).

\bibitem[{\citenamefont{Weinert and Scheffler}(1986)}]{WeinerMSF86}
\bibinfo{author}{\bibfnamefont{C.~M.} \bibnamefont{Weinert}} \bibnamefont{and}
  \bibinfo{author}{\bibfnamefont{M.}~\bibnamefont{Scheffler}},
  \bibinfo{journal}{Mater. Sci. Forum} \textbf{\bibinfo{volume}{10-12}},
  \bibinfo{pages}{25} (\bibinfo{year}{1986}).

\bibitem[{\citenamefont{Wang et~al.}(2000)\citenamefont{Wang, Chaka, and
  Scheffler}}]{WangPRL2000}
\bibinfo{author}{\bibfnamefont{X.-G.} \bibnamefont{Wang}},
  \bibinfo{author}{\bibfnamefont{A.}~\bibnamefont{Chaka}}, \bibnamefont{and}
  \bibinfo{author}{\bibfnamefont{M.}~\bibnamefont{Scheffler}},
  \bibinfo{journal}{Phys. Rev. Lett.} \textbf{\bibinfo{volume}{84}},
  \bibinfo{pages}{3650} (\bibinfo{year}{2000}).

\bibitem[{\citenamefont{Reuter and Scheffler}(2002)}]{ReuterPRB2002}
\bibinfo{author}{\bibfnamefont{K.}~\bibnamefont{Reuter}} \bibnamefont{and}
  \bibinfo{author}{\bibfnamefont{M.}~\bibnamefont{Scheffler}},
  \bibinfo{journal}{Phys. Rev. B} \textbf{\bibinfo{volume}{65}},
  \bibinfo{pages}{035406} (\bibinfo{year}{2002}).

\bibitem[{\citenamefont{Li et~al.}(2003)\citenamefont{Li, Stampfl, and
  Scheffler}}]{LiPRB2003}
\bibinfo{author}{\bibfnamefont{W.-X.} \bibnamefont{Li}},
  \bibinfo{author}{\bibfnamefont{C.}~\bibnamefont{Stampfl}}, \bibnamefont{and}
  \bibinfo{author}{\bibfnamefont{M.}~\bibnamefont{Scheffler}},
  \bibinfo{journal}{Phys. Rev. B} \textbf{\bibinfo{volume}{67}},
  \bibinfo{pages}{045408} (\bibinfo{year}{2003}).

\bibitem[{\citenamefont{Reuter and Scheffler}(2001)}]{ReuterRuO2}
\bibinfo{author}{\bibfnamefont{K.}~\bibnamefont{Reuter}} \bibnamefont{and}
  \bibinfo{author}{\bibfnamefont{M.}~\bibnamefont{Scheffler}},
  \bibinfo{journal}{Phys. Rev. B} \textbf{\bibinfo{volume}{65}},
  \bibinfo{pages}{035406} (\bibinfo{year}{2001}).

\bibitem[{\citenamefont{Stull and Prophet}(1971)}]{thermotables}
\bibinfo{author}{\bibfnamefont{D.~R.} \bibnamefont{Stull}} \bibnamefont{and}
  \bibinfo{author}{\bibfnamefont{H.}~\bibnamefont{Prophet}},
  \emph{\bibinfo{title}{JANAF Thermochemical Tables, 2nd Ed.}}
  (\bibinfo{publisher}{U.S. National Bureau of Standards},
  \bibinfo{address}{Washington, DC}, \bibinfo{year}{1971}).

\bibitem[{\citenamefont{Tersoff}(1995)}]{TersoffPRL1995}
\bibinfo{author}{\bibfnamefont{J.}~\bibnamefont{Tersoff}},
  \bibinfo{journal}{Phys. Rev. Lett.} \textbf{\bibinfo{volume}{74}},
  \bibinfo{pages}{434} (\bibinfo{year}{1995}).

\bibitem[{\citenamefont{Neugebauer and Scheffler}(1993)}]{NeugebauerPRL93}
\bibinfo{author}{\bibfnamefont{J.}~\bibnamefont{Neugebauer}} \bibnamefont{and}
  \bibinfo{author}{\bibfnamefont{M.}~\bibnamefont{Scheffler}},
  \bibinfo{journal}{Phys. Rev. Lett.} \textbf{\bibinfo{volume}{71}},
  \bibinfo{pages}{577} (\bibinfo{year}{1993}).

\bibitem[{\citenamefont{Pleth~Nielsen et~al.}(1993)\citenamefont{Pleth~Nielsen,
  Besenbacher, Stensgaard, Laegsgaard, Engdahl, Stoltze, Jacobsen, and
  N\o{}rskov}}]{NielsenPRL93}
\bibinfo{author}{\bibfnamefont{L.}~\bibnamefont{Pleth~Nielsen}},
  \bibinfo{author}{\bibfnamefont{F.}~\bibnamefont{Besenbacher}},
  \bibinfo{author}{\bibfnamefont{I.}~\bibnamefont{Stensgaard}},
  \bibinfo{author}{\bibfnamefont{E.}~\bibnamefont{Laegsgaard}},
  \bibinfo{author}{\bibfnamefont{C.}~\bibnamefont{Engdahl}},
  \bibinfo{author}{\bibfnamefont{P.}~\bibnamefont{Stoltze}},
  \bibinfo{author}{\bibfnamefont{K.~W.} \bibnamefont{Jacobsen}},
  \bibnamefont{and} \bibinfo{author}{\bibfnamefont{J.~K.}
  \bibnamefont{N\o{}rskov}}, \bibinfo{journal}{Phys. Rev. Lett.}
  \textbf{\bibinfo{volume}{71}}, \bibinfo{pages}{754} (\bibinfo{year}{1993}).

\bibitem[{\citenamefont{R\o{}der et~al.}(1993)\citenamefont{R\o{}der, Shuster,
  Brune, and Kern}}]{RoderPRL93}
\bibinfo{author}{\bibfnamefont{H.}~\bibnamefont{R\o{}der}},
  \bibinfo{author}{\bibfnamefont{R.}~\bibnamefont{Shuster}},
  \bibinfo{author}{\bibfnamefont{H.}~\bibnamefont{Brune}}, \bibnamefont{and}
  \bibinfo{author}{\bibfnamefont{K.}~\bibnamefont{Kern}},
  \bibinfo{journal}{Phys. Rev. Lett.} \textbf{\bibinfo{volume}{71}},
  \bibinfo{pages}{2086} (\bibinfo{year}{1993}).

\bibitem[{\citenamefont{Oppo et~al.}(1993)\citenamefont{Oppo, Fiorentini, and
  Scheffler}}]{OppoPRL93}
\bibinfo{author}{\bibfnamefont{S.}~\bibnamefont{Oppo}},
  \bibinfo{author}{\bibfnamefont{V.}~\bibnamefont{Fiorentini}},
  \bibnamefont{and}
  \bibinfo{author}{\bibfnamefont{M.}~\bibnamefont{Scheffler}},
  \bibinfo{journal}{Phys. Rev. Lett.} \textbf{\bibinfo{volume}{71}},
  \bibinfo{pages}{2437} (\bibinfo{year}{1993}).

\bibitem[{\citenamefont{Boer et~al.}(1988)\citenamefont{Boer, Boom, Mattens,
  Miederna, and Niessen}}]{BoerBOOK}
\bibinfo{author}{\bibfnamefont{F.~R.~D.} \bibnamefont{Boer}},
  \bibinfo{author}{\bibfnamefont{R.}~\bibnamefont{Boom}},
  \bibinfo{author}{\bibfnamefont{W.~C.~M.} \bibnamefont{Mattens}},
  \bibinfo{author}{\bibfnamefont{A.~R.} \bibnamefont{Miederna}},
  \bibnamefont{and} \bibinfo{author}{\bibfnamefont{A.~K.}
  \bibnamefont{Niessen}}, \emph{\bibinfo{title}{Cohesion in metals}}
  (\bibinfo{publisher}{North-Holland, Amsterdam}, \bibinfo{year}{1988}).

\bibitem[{\citenamefont{Lindgren et~al.}(1984)\citenamefont{Lindgren,
  Walld\'en, Rundgren, and Westrin}}]{Lindgren84}
\bibinfo{author}{\bibfnamefont{S.~A.} \bibnamefont{Lindgren}},
  \bibinfo{author}{\bibfnamefont{L.}~\bibnamefont{Walld\'en}},
  \bibinfo{author}{\bibfnamefont{J.}~\bibnamefont{Rundgren}}, \bibnamefont{and}
  \bibinfo{author}{\bibfnamefont{P.}~\bibnamefont{Westrin}},
  \bibinfo{journal}{Phys. Rev. B} \textbf{\bibinfo{volume}{29}},
  \bibinfo{pages}{576} (\bibinfo{year}{1984}).

\bibitem[{\citenamefont{Christensen et~al.}(1997)\citenamefont{Christensen,
  Ruban, Stoltze, Jacobsen, Skiver, N\o{}rskov, and
  Besenbacher}}]{NorskovSurfAll}
\bibinfo{author}{\bibfnamefont{A.}~\bibnamefont{Christensen}},
  \bibinfo{author}{\bibfnamefont{A.~V.} \bibnamefont{Ruban}},
  \bibinfo{author}{\bibfnamefont{P.}~\bibnamefont{Stoltze}},
  \bibinfo{author}{\bibfnamefont{K.~W.} \bibnamefont{Jacobsen}},
  \bibinfo{author}{\bibfnamefont{H.~K.} \bibnamefont{Skiver}},
  \bibinfo{author}{\bibfnamefont{J.~K.} \bibnamefont{N\o{}rskov}},
  \bibnamefont{and}
  \bibinfo{author}{\bibfnamefont{F.}~\bibnamefont{Besenbacher}},
  \bibinfo{journal}{Phys. Rev. B} \textbf{\bibinfo{volume}{56}},
  \bibinfo{pages}{5822} (\bibinfo{year}{1997}).

\bibitem[{\citenamefont{Maurel et~al.}(2005)\citenamefont{Maurel, Abel, Koudia,
  Bocquet, and Porte}}]{MaurelSS2005}
\bibinfo{author}{\bibfnamefont{C.}~\bibnamefont{Maurel}},
  \bibinfo{author}{\bibfnamefont{M.}~\bibnamefont{Abel}},
  \bibinfo{author}{\bibfnamefont{M.}~\bibnamefont{Koudia}},
  \bibinfo{author}{\bibfnamefont{F.}~\bibnamefont{Bocquet}}, \bibnamefont{and}
  \bibinfo{author}{\bibfnamefont{L.}~\bibnamefont{Porte}},
  \bibinfo{journal}{Surf. Sci.} \textbf{\bibinfo{volume}{596}},
  \bibinfo{pages}{45} (\bibinfo{year}{2005}).

\bibitem[{\citenamefont{Bocquet et~al.}(2005)\citenamefont{Bocquet, Maurel,
  Roussel, Abel, Koudia, and Porte}}]{BocquetPRB2005}
\bibinfo{author}{\bibfnamefont{F.}~\bibnamefont{Bocquet}},
  \bibinfo{author}{\bibfnamefont{C.}~\bibnamefont{Maurel}},
  \bibinfo{author}{\bibfnamefont{J.}~\bibnamefont{Roussel}},
  \bibinfo{author}{\bibfnamefont{M.}~\bibnamefont{Abel}},
  \bibinfo{author}{\bibfnamefont{M.}~\bibnamefont{Koudia}}, \bibnamefont{and}
  \bibinfo{author}{\bibfnamefont{L.}~\bibnamefont{Porte}},
  \bibinfo{journal}{Phys. Rev. B} \textbf{\bibinfo{volume}{71}},
  \bibinfo{pages}{075405} (\bibinfo{year}{2005}).

\bibitem[{\citenamefont{Kitchin et~al.}(2008)\citenamefont{Kitchin, Reuter, and
  Scheffler}}]{KitchinPRB2008}
\bibinfo{author}{\bibfnamefont{J.~R.} \bibnamefont{Kitchin}},
  \bibinfo{author}{\bibfnamefont{K.}~\bibnamefont{Reuter}}, \bibnamefont{and}
  \bibinfo{author}{\bibfnamefont{M.}~\bibnamefont{Scheffler}},
  \bibinfo{journal}{Phys. Rev. B} \textbf{\bibinfo{volume}{77}},
  \bibinfo{pages}{075437} (\bibinfo{year}{2008}).

\bibitem[{\citenamefont{Han et~al.}(2005)\citenamefont{Han, der Ven, Ceder, and
  Hwang}}]{HanPRB2005}
\bibinfo{author}{\bibfnamefont{B.~C.} \bibnamefont{Han}},
  \bibinfo{author}{\bibfnamefont{A.~V.} \bibnamefont{der Ven}},
  \bibinfo{author}{\bibfnamefont{G.}~\bibnamefont{Ceder}}, \bibnamefont{and}
  \bibinfo{author}{\bibfnamefont{B.-J.} \bibnamefont{Hwang}},
  \bibinfo{journal}{Phys. Rev. B} \textbf{\bibinfo{volume}{72}},
  \bibinfo{pages}{205409} (\bibinfo{year}{2005}).

\bibitem[{\citenamefont{Seriani and Mittendorferl}(2008)}]{SerianiJPCM2008}
\bibinfo{author}{\bibfnamefont{N.}~\bibnamefont{Seriani}} \bibnamefont{and}
  \bibinfo{author}{\bibfnamefont{F.}~\bibnamefont{Mittendorferl}},
  \bibinfo{journal}{J. Phys.: Condens. Matter} \textbf{\bibinfo{volume}{20}},
  \bibinfo{pages}{184023} (\bibinfo{year}{2008}).

\bibitem[{\citenamefont{Soon et~al.}(2005)\citenamefont{Soon, Todorova, Delley,
  and Stampfl}}]{SoonPRB2006}
\bibinfo{author}{\bibfnamefont{A.}~\bibnamefont{Soon}},
  \bibinfo{author}{\bibfnamefont{M.}~\bibnamefont{Todorova}},
  \bibinfo{author}{\bibfnamefont{B.}~\bibnamefont{Delley}}, \bibnamefont{and}
  \bibinfo{author}{\bibfnamefont{C.}~\bibnamefont{Stampfl}},
  \bibinfo{journal}{Phys. Rev. B} \textbf{\bibinfo{volume}{73}},
  \bibinfo{pages}{165424} (\bibinfo{year}{2005}).

\bibitem[{\citenamefont{Kim et~al.}(2003)\citenamefont{Kim, Rodriguez, Hanson,
  Frenkel, and Lee}}]{KimJACS03}
\bibinfo{author}{\bibfnamefont{J.~Y.} \bibnamefont{Kim}},
  \bibinfo{author}{\bibfnamefont{J.~A.} \bibnamefont{Rodriguez}},
  \bibinfo{author}{\bibfnamefont{J.~C.} \bibnamefont{Hanson}},
  \bibinfo{author}{\bibfnamefont{A.~I.} \bibnamefont{Frenkel}},
  \bibnamefont{and} \bibinfo{author}{\bibfnamefont{P.}~\bibnamefont{Lee}},
  \bibinfo{journal}{J. Am. Chem. Soc.} \textbf{\bibinfo{volume}{125}},
  \bibinfo{pages}{10684} (\bibinfo{year}{2003}).

\bibitem[{CRC(1995)}]{CRC}
\emph{\bibinfo{title}{CRC Handbook of Chemistry and Physics}}
  (\bibinfo{publisher}{CRC Press, Boca Raton FL}, \bibinfo{year}{1995}).

\bibitem[{\citenamefont{Jiang et~al.}()\citenamefont{Jiang, Gomez-Abal, Li,
  Rinke, and Scheffler}}]{JiangTBP}
\bibinfo{author}{\bibfnamefont{H.}~\bibnamefont{Jiang}},
  \bibinfo{author}{\bibfnamefont{R.}~\bibnamefont{Gomez-Abal}},
  \bibinfo{author}{\bibfnamefont{X.}~\bibnamefont{Li}},
  \bibinfo{author}{\bibfnamefont{P.}~\bibnamefont{Rinke}}, \bibnamefont{and}
  \bibinfo{author}{\bibfnamefont{M.}~\bibnamefont{Scheffler}},
  \bibinfo{note}{to be published}.

\bibitem[{\citenamefont{Hu et~al.}(2007)\citenamefont{Hu, Reuter, and
  Scheffler}}]{HuReutPRL2007}
\bibinfo{author}{\bibfnamefont{Q.-M.} \bibnamefont{Hu}},
  \bibinfo{author}{\bibfnamefont{K.}~\bibnamefont{Reuter}}, \bibnamefont{and}
  \bibinfo{author}{\bibfnamefont{M.}~\bibnamefont{Scheffler}},
  \bibinfo{journal}{Phys. Rev. Lett.} \textbf{\bibinfo{volume}{98}},
  \bibinfo{pages}{176103} (\bibinfo{year}{2007}).

\bibitem[{\citenamefont{Klikovits et~al.}(2007)\citenamefont{Klikovits,
  Napetschnig, Schmid, Seriani, Dubay, Kresse, and Varga}}]{KlikovitsPRB2007}
\bibinfo{author}{\bibfnamefont{J.}~\bibnamefont{Klikovits}},
  \bibinfo{author}{\bibfnamefont{E.}~\bibnamefont{Napetschnig}},
  \bibinfo{author}{\bibfnamefont{M.}~\bibnamefont{Schmid}},
  \bibinfo{author}{\bibfnamefont{N.}~\bibnamefont{Seriani}},
  \bibinfo{author}{\bibfnamefont{O.}~\bibnamefont{Dubay}},
  \bibinfo{author}{\bibfnamefont{G.}~\bibnamefont{Kresse}}, \bibnamefont{and}
  \bibinfo{author}{\bibfnamefont{P.}~\bibnamefont{Varga}},
  \bibinfo{journal}{Phys. Rev. B} \textbf{\bibinfo{volume}{76}},
  \bibinfo{pages}{045405} (\bibinfo{year}{2007}).

\bibitem[{\citenamefont{Zafeiratos et~al.}()\citenamefont{Zafeiratos,
  H\"{a}vecker, Teschner, Vass, Schn\"{o}rch, Girgsdies, Hansen, Knop-Gericke,
  Schl\"{o}ögl, and Bukhtiyarov}}]{SpirosUNP}
\bibinfo{author}{\bibfnamefont{S.}~\bibnamefont{Zafeiratos}},
  \bibinfo{author}{\bibfnamefont{M.}~\bibnamefont{H\"{a}vecker}},
  \bibinfo{author}{\bibfnamefont{D.}~\bibnamefont{Teschner}},
  \bibinfo{author}{\bibfnamefont{E.}~\bibnamefont{Vass}},
  \bibinfo{author}{\bibfnamefont{P.}~\bibnamefont{Schn\"{o}rch}},
  \bibinfo{author}{\bibfnamefont{F.}~\bibnamefont{Girgsdies}},
  \bibinfo{author}{\bibfnamefont{T.}~\bibnamefont{Hansen}},
  \bibinfo{author}{\bibfnamefont{A.}~\bibnamefont{Knop-Gericke}},
  \bibinfo{author}{\bibfnamefont{R.}~\bibnamefont{Schl\"{o}ögl}},
  \bibnamefont{and}
  \bibinfo{author}{\bibfnamefont{V.}~\bibnamefont{Bukhtiyarov}},
  \bibinfo{note}{unpublished}.

\end{thebibliography}

\end{document}